\documentclass[manuscript]{emulateapj} 
\usepackage{natbib}
\usepackage{graphicx} 
\usepackage{txfonts}
\usepackage{textcomp}
\usepackage{url}
\usepackage{hyperref}
\hypersetup{
    colorlinks=false,
    pdfborder={0 0 0},
}
\makeatletter 

\newcommand{\msun}{{M$_{\odot}~$}}

\slugcomment{Accepted for publication in ApJ}
\shorttitle{Stellar Kinematics of $z\sim2$ galaxies} 
\shortauthors{van de Sande et al.}  

\begin{document}

\title{Stellar Kinematics of $z\sim2$ galaxies and the inside-out growth of quiescent galaxies \altaffilmark{1,}\altaffilmark{2}}

\author{Jesse van de Sande\altaffilmark{3}, 
Mariska Kriek\altaffilmark{4}, 
Marijn Franx\altaffilmark{3}, 
Pieter G. van Dokkum\altaffilmark{5}, 
Rachel Bezanson\altaffilmark{5}, 
Rychard J. Bouwens\altaffilmark{3}, 
Ryan F. Quadri \altaffilmark{6}, 
Hans-Walter Rix \altaffilmark{7},
Rosalind E. Skelton \altaffilmark{5}
} 

\altaffiltext{1}{Based on X-Shooter-VLT observations collected at the
European Southern Observatory, Paranal, Chile.}

\altaffiltext{2} {Based on observations with the NASA/ESA
\textit{Hubble Space Telescope (HST)}, obtained at the Space Telescope
Science Institute, which is operated by AURA, Inc., under NASA
contract NAS 5-26555.}

\altaffiltext{3}{Leiden Observatory, Leiden University, P.O.\ Box
9513, 2300 RA Leiden, The Netherlands.}

\altaffiltext{4}{Astronomy Department, Berkeley,
Hearst Field Annex, Berkeley, CA 94720-3411, USA}

\altaffiltext{5}{Department of Astronomy, Yale University, P.O. \ Box
208101, New Haven, CT 06520-8101, USA.}

\altaffiltext{6}{Carnegie Observatories, Pasadena, CA 91101, USA}

\altaffiltext{7}{Max Planck Institute for Astronomy, K\"onigstuhl 17, D-
69117 Heidelberg, Germany}


\begin{abstract}
Using stellar kinematics measurements, we investigate the growth of
massive, quiescent galaxies from $z\sim$2 to today. We present
X-Shooter spectra from the UV to NIR and dynamical mass measurements
of five quiescent massive ($>10^{11}$ \msun) galaxies at $z\sim2$. This
triples the sample of $z>1.5$ galaxies with well constrained
($\delta\sigma<100$km s$^{-1}$) velocity dispersion measurements. From
spectral population synthesis modeling we find that these galaxies
have stellar ages that range from 0.5-2 Gyr, with no signs of ongoing
star formation. We measure velocity dispersions (290-450 km s$^{-1}$)
from stellar absorption lines and find that they are 1.6-2.1 times
higher than those of galaxies in the Sloan Digital Sky Survey at the
same mass. Sizes are measured using GALFIT from Hubble Space Telescope
Wide Field Camera 3 $H_{160}$ and UDS K-band images. The dynamical
masses correspond well to the spectral energy distribution based
stellar masses, with dynamical masses that are $\sim15\%$ higher. We
find that $M_*/M_{{dyn}}$ may decrease slightly with time, which could
reflect the increase of the dark matter fraction within an increasing
effective radius.  We combine different stellar kinematic studies from
the literature, and examine the structural evolution from $z\sim2$ to
$z\sim0$: we confirm that at fixed dynamical mass, the effective
radius increases by a factor of $\sim2.8$, and the velocity dispersion
decreases by a factor of $\sim1.7$. The mass density within one
effective radius decreases by a factor of $\sim20$, while within a
fixed physical radius (1 kpc) it decreases only mildly (factor of
$\sim2)$. When we allow for an evolving mass limit by selecting a
population of galaxies at fixed number density, a stronger size growth
with time is found (factor of $\sim4)$, velocity dispersion decreases
by a factor of $\sim1.4$, and interestingly, the mass density within 1
kpc is consistent with no evolution. This finding suggests that
massive quiescent galaxies at $z\sim2$ grow inside-out, consistent
with the expectations from minor mergers.
\end{abstract}


\section{Introduction}
\label{sec:introduction} 
Recent studies have shown that a considerable fraction of massive
galaxies at $1.5 < z < 2.5$ have quiescent stellar populations
(e.g.,\citealt{labbe2005}; \citealt{kriek2006};
\citealt{williams2009}). Among the most massive galaxies ($M_{*} >
10^{11}$ \msun ) approximately 40\% are no longer forming stars
(e.g.,\citealt{whitaker2011}; \citealt{brammer2011}). Surprisingly,
these massive quiescent galaxies have been found to be extremely
compact (e.g.,\citealt{daddi2005}; \citealt{trujillo2006};
\citealt{vandokkum2008}; \citealt{franx2008}; \citealt{vanderwel2008};
and numerous others), compared to their likely present-day
counterparts.

Searches for ultra-dense low-redshift counterparts by
\citet{trujillo2009} and \citet{taylor2010a} found only a handful of
compact sources at $z\sim0$, which have relatively young stellar
populations (\citealt{trujillo2009}; \citealt{ferremateu2012}).  The
dearth of massive, old compact objects at low redshift implies that
massive galaxies must have undergone severe structural evolution in
size.

Errors in the size estimates have been invoked as a possible
explanation for the compactness of massive high-redshift galaxies.
Initial concerns that the size may have been underestimated, due to an
envelope of low surface brightness light, have been addressed with
deep Hubble Space Telescope Wide Field Camera 3 (HST WFC3) imaging
(\citealt{szomoru2010}; \citeyear{szomoru2012}), and by stacking
results (e.g \citealt{vanderwel2008}; \citealt{cassata2010};
\citealt{vandokkum2008}, \citeyear{vandokkum2010}). The light could
also be more concentrated due to the presence of active galactic
nuclei (AGNs) in these galaxies. However, spectra of subsamples of
these galaxies have shown that the light is dominated by evolved
stellar populations, not AGNs (\citeauthor{kriek2006}
\citeyear{kriek2006}, \citeyear{kriek2009a}; \citealt{vandesande2011};
\citealt{onodera2012}).

The question of whether stellar masses are accurate out to $z\sim 2$
remains, however, a serious concern: an overestimate in stellar mass
would bring the galaxies closer to the $z\sim0$ mass-size relation. To
date, basically all (stellar) masses have been derived by fitting the
spectral energy distributions (SEDs). This method suffers from many
systematic uncertainties in stellar population synthesis (SPS) models 
(e.g \citealt{conroy2009}; \citealt{muzzin2009}) and is essentially
untested at $z > 1.5$.


\begin{deluxetable*}{c c c c c c c c c c c c}
\tabletypesize{\scriptsize}
\tablewidth{0pt}
\tablecaption{Photometric Properties}
\tablehead{
\colhead{Catalog} & \colhead{ID} & \colhead{$J_{\rm aper}$}  & \colhead{$H_{\rm aper}$} & \colhead{$K_{\rm aper}$} & \colhead{$K_{\rm tot}$}  & \colhead{$(U-V)_{\rm~rf}$}  & \colhead{$(V-J)_{\rm~rf}$} & \colhead{24 $\mu \rm{m}$}  & \colhead{SFR$_{24\mu m}$} \\
& & &  &  &   &  & & \colhead{ $[\mu\rm{Jy}]$}  & \colhead{\msun\ yr$^{-1}$} }\\
\startdata
NMBS-COS  & 7447  &    21.09  &    20.72  &    20.63  &    19.64  &     1.20  &  0.31 & $\lesssim 18$ & $\lesssim 13$\\
NMBS-COS  &18265 &    22.67  &    20.85  &    20.61  &    19.62  &     1.72  &   0.85 & $\lesssim 18$ & $\lesssim 15$\\
NMBS-COS  & 7865  &    22.75  &    21.51  &    21.07  &    20.02  &     1.89  &   0.90 & $\lesssim 18$ & $\lesssim 19$\\
   UDS  &19627 &    21.40  &    20.91  &    20.65  &    20.19  &     1.37  &  0.71  & $\lesssim 30$ & $\lesssim 29$\\
   UDS  &29410  &    20.59  &    20.18  &    19.81  &    19.36  &     1.62  &  0.96  & $232 \pm 15$ & $241\pm16$ \\
\enddata
\tablecomments{Aperture and total magnitudes for our targets. Aperture
magnitudes have been measured in fixed 1.5 arcsec diameter aperture
for targets in NMBS-I, while the targets in UDS have 1.75 arcsec
diameter apertures. Rest-frame colors have been derived from the
spectra in Johnson $U$, $V$, and $2MASS J$ filters. For the 24 $\mu$m
fluxes we provide 3-$\sigma$ upper-limits of 18 $\mu$Jy for the
galaxies in NMBS-COSMOS \citep{whitaker2012}, and 30 $\mu$Jy for
UDS-19627 \citep{toft2012}, as these galaxies are not detected with
MIPS. UDS-29410 has a strong MIPS detection, which is likely due to an
obscured AGN.}
\label{tab:phot_sample}
\end{deluxetable*}

Direct stellar kinematic mass measurements, which do not suffer from
these uncertainties, can be derived by measuring the galaxy's velocity
dispersion and the shape and extent of its luminosity profile,
i.e.,the S\'ersic $n$ parameter and effective radius. In particular,
for low-redshift galaxies in the Sloan Digital Sky Survey (SDSS),
\citet{taylor2010b} showed that stellar mass is a very good predictor
of dynamical mass, but only when non-homology of luminosity profile is
properly accounted for using a S\'ersic-dependent virial factor
(e.g.,\citealt{cappellari2006}). Although dynamical measurements of
massive galaxies are common at low redshift, spectroscopic studies
become much more difficult at higher redshift as the bulk of the
light, and stellar absorption features used to measure kinematics,
shift redward into the near-infrared (NIR) (e.g.,\citealt{kriek2009a};
\citealt{vandokkum2009a}).

New technology such as the new red arm of the LRIS spectrograph at
Keck (working beyond 1$\micron$) makes it possible to measure velocity
dispersions up to $z\sim1.5$ (\citealt{newman2010},
\citealt{bezanson2012}). Deep NIR spectroscopy is, however, required
to push stellar kinematic studies to even higher redshift. From a
$\sim$29\,hr spectrum of an ultra-compact galaxy at $z=2.2$ obtained
with Gemini Near-IR Spectrograph \citep{kriek2009a},
\citet{vandokkum2009a} found a high, though uncertain, velocity
dispersion of $\sigma =
510^{+165}_{-95}\,$km$~$s$^{-1}$. \citet{onodera2012} used the MOIRCS
on the Subaru telescope to observe the rest-frame optical spectrum of
a less-compact, passive, ultra-massive galaxy at $z=1.82$, but the low
spectral resolution and signal-to-noise ratio (S/N) severely limited the accuracy
of their velocity dispersion: $\sigma = 270 {\pm
105}\,$km$~$s$^{-1}$. X-Shooter (\citealt{dodorico2006};
\citealt{vernet2011}), the new ultraviolet (UV) to NIR spectrograph at
the Very Large Telescope (VLT), can provide the required S/N and
resolution. The capabilities of X-Shooter for this kind of
measurements were demonstrated in \citet{vandesande2011}, who found
$294 \pm 51\,$km$~$s$^{-1}$ for a massive quiescent galaxy at
z=1.8. \citet{toft2012} also use X-Shooter and present a dynamical
measurement of a galaxy at redshift $z=2.04$ with similar
results. Taken all together, these results indicate that the dynamical
and stellar masses are consistent with $z\sim0$.  With the small
number of measurements beyond $z>1.5$, however, the sample is still
too small to draw any firm conclusions on whether the stellar masses
are truly reliable.

Here we present a sample of five massive quiescent galaxies with high
signal-to-noise (S/N), medium-resolution, UV to NIR spectra at
$1.4<z<2.1$ observed with X-Shooter on the VLT. The main goal of this
paper will be to test if the stellar mass measurements at high
redshift are reliable.

The paper is organized as follows. In \textsection \ref{sec:data} we
present our sample of high-redshift galaxies, the photometric and
spectroscopic data, and describe our data reduction. In \textsection
\ref{sec:properties} we determine structural properties and stellar
populations, and derive stellar and dynamical masses. We complement
our results with stellar kinematic results from other studies at low
and high redshift in \textsection \ref{sec:compilation}. In
\textsection \ref{sec:mass_comparison} we compare our dynamical to the
stellar masses. In \textsection \ref{sec:evolution} we study the
structural evolution of high-redshift quiescent massive galaxies. In
\textsection \ref{sec:inside_out} we compare our results with previous
measurements and hydrodynamical simulations. Finally, in \textsection
\ref{sec:conclusion} we summarize our results and
conclusions. Throughout the paper we assume a $\Lambda$CDM cosmology
with $\Omega_\mathrm{m}$=0.3, $\Omega_{\Lambda}=0.7$, and $H_{0}=70$
km s$^{-1}$ Mpc$^{-1}$. All broadband data are given in the AB-based
photometric system.


\begin{figure*}[ht] \epsscale{1.1}
\plotone{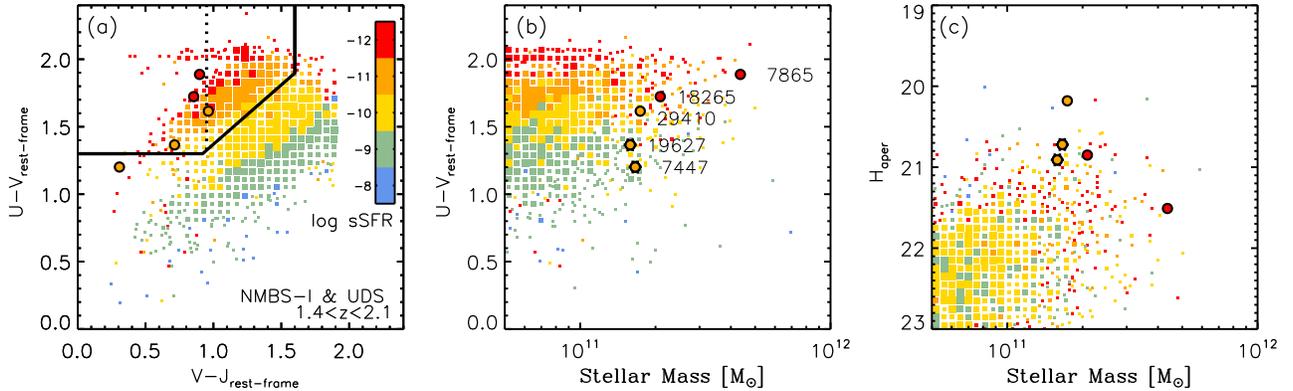}
\caption{Comparison of our spectroscopic sample to the full population
at similar redshift. Symbol size of the squares represent the density
of galaxies from the NMBS-I and UDS at $1.4<z<2.1$ with mass
$>10^{10.5} M_{\odot}$. (a) Rest-frame U-V and V-J
colors. Color coding is based upon the sSFR derived from SED fitting,
red colors indicates low sSFR (quiescent), and blue colors indicate high
sSFR(star-forming). Galaxies in the top left region, as marked by the
black line, all have low sSFR rates. This region is therefore often
used to select quiescent galaxies at high redshift
\citep{williams2009}. All but one of our galaxies fall within this
region, but their sSFR indicate that they are all have quiescent
stellar populations. The vertical dotted line discriminates between
young post-starburst like (left) vs. old quiescent (right) as
indicated by \citet{whitaker2012}. The strong Balmer absorption lines
spectroscopically confirm the young ages of this
sample. (b) Rest-frame U-V vs. stellar mass. At
fixed mass, we find that most of our galaxies have similar colors to
the entire population, except for NMBS-COS7447 and UDS-19627 on the
blue side. (c) $H$-band aperture magnitude vs.
stellar mass. It is clear that our sample was selected on magnitude,
and at fixed mass they are among the brightest galaxies, consistent
with their post-starburst nature.
}
\label{fig:selection_effects_sp}
\end{figure*}
%


\begin{deluxetable*}{c c c c c c c c c l}[!ht]
\tabletypesize{\scriptsize}
\tablewidth{0pt}
\tablecaption{Targets and Observations}
\tablehead{
\colhead{Catalog} & \colhead{ID} & \colhead{R.A.} & \colhead{Dec.} & \colhead{Exp. Time} & \colhead{Slit Size NIR} & \colhead{S/N $J$} & \colhead{S/N $H$}& \colhead{S/N $_{{4020<\lambda \AA<7000}}$} & \colhead{Telluric Standard Star} \\
 &  & &  & \colhead{[min]} & \colhead{[arcsec]} & \colhead{[\AA$^{-1}$]} & \colhead{[\AA$^{-1}$]}  & \colhead{[\AA$^{-1}$]} & \colhead{$Hipparcos~ID$} }\\
\startdata
\vspace{0.15cm}
NMBS-COS  & 7447  &10:00: 6.96  & 2:17:33.77  &  120  &      0''.9  &    4.98  &    8.48  &    6.31  &    050307, 000349 \\
\vspace{0.15cm}
NMBS-COS  &18265  &10:00:40.83  & 2:28:52.15  &   90  &      0''.9  &    3.37  &    6.99  &    4.18  &    050684, 000349 \\
NMBS-COS  & 7865  &10:00:17.73  & 2:17:52.75  &  434  &      0''.9  &    1.64  &    5.86  &    4.12  &     049704, 057126, 046054, \\ 
 \vspace{0.05cm}
& & & & & & & & & 040217, 059987 \\
UDS  &19627  & 2:18:17.06  &-5:21:38.83  &  300  &0''.6, 0''.9  &    3.80  &    7.94  &    5.90  &       012377, 114656, 008352, \\ 
\vspace{0.05cm}
& & & & & & & & & 000328, 015389 \\
UDS  &29410  & 2:17:51.22  &-5:16:21.84  &  120  &      0''.9  &    3.75  &    7.09  &    4.35  &   012377
\enddata
\tablecomments{R.A. and Dec. are given in the J2000 coordinate
system. Exposure times are given for the NIR arm, the UVB and VIS arms
had slightly shorter exposure times due to the longer read out. Except
for UDS19627, all targets were observed with a 0.9" slit in the
NIR. S/N ratios have been determined from comparing the residual of
the velocity dispersion fit to the flux, and are given for the $J-$
and $H-$band as well as for the region in which we determine the
velocity dispersion. Last column gives the Tellurics Standard Stars
from the Hipparcos catalog that were observed before and after each
target.}
\label{tab:sample_cat}
\end{deluxetable*}

\begin{figure*}
 \epsscale{1.1}
\plotone{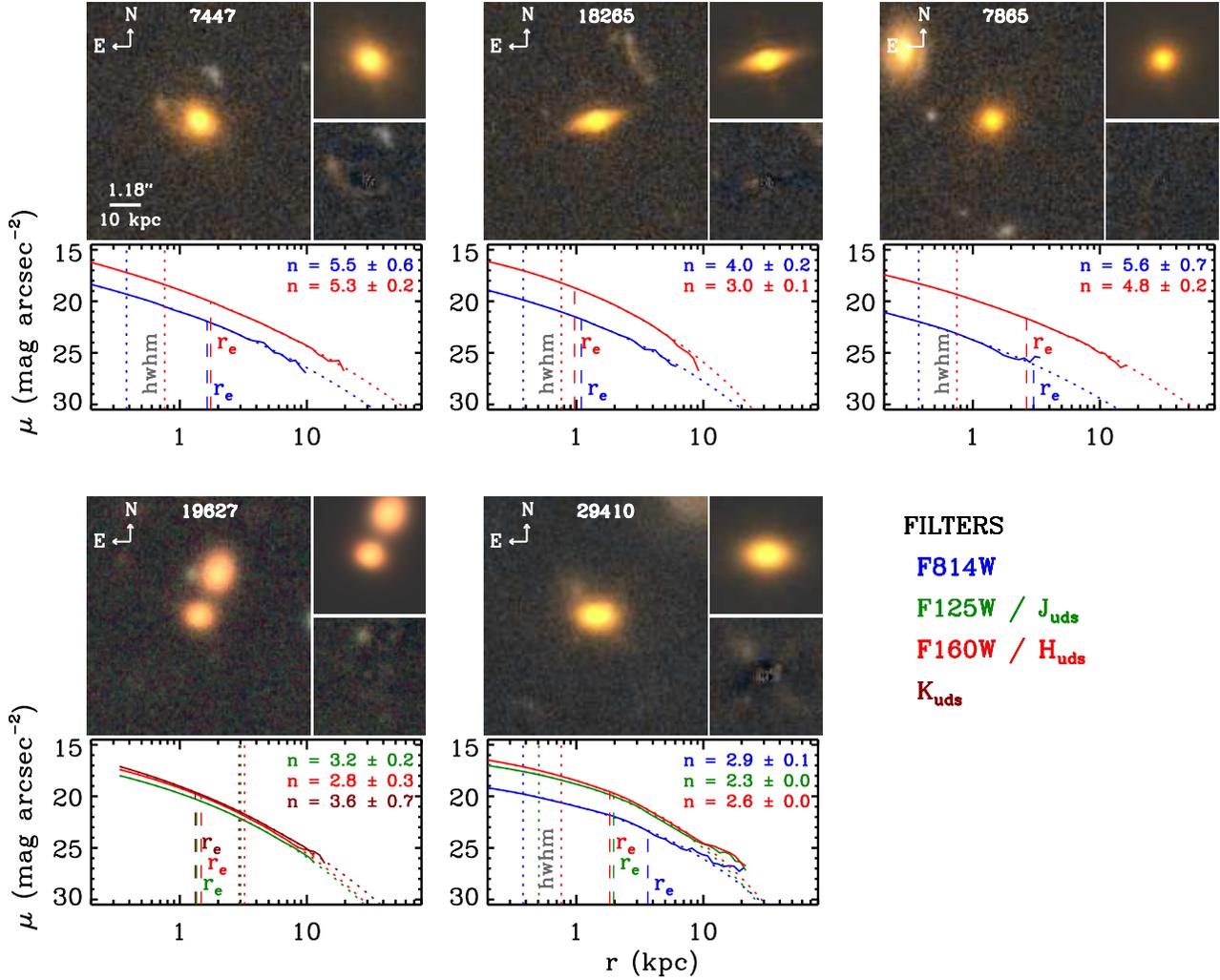}
\caption{Color Images of our five spectroscopic targets. Except for
UDS-19627, all galaxies have available HST-WFC3 Imaging. For each
target we show the composite color image on the left side, the best
S\'ersic model from GALFIT and the residual after we subtract the
best-fitting model from the original image on the right side. The
lower panel shows the intrinsic surface brightness profile with all
available bands. Different colors show the different filters, as
indicated on the bottom right. Vertical dashed lines show the
effective radii for each profile, while the dotted lines shows the
FWHM/2 of the PSF.  We find color gradients, such that the redder
bands have smaller effective radii, for all galaxies but
NMBS-COS7447. For that case, the sizes are similar within the errors,
but this could be caused by the extra flux of the red arc-like feature
in the southeast.
}
\label{fig:sizes}
\end{figure*}


\begin{figure*}
\epsscale{1.1}
\plotone{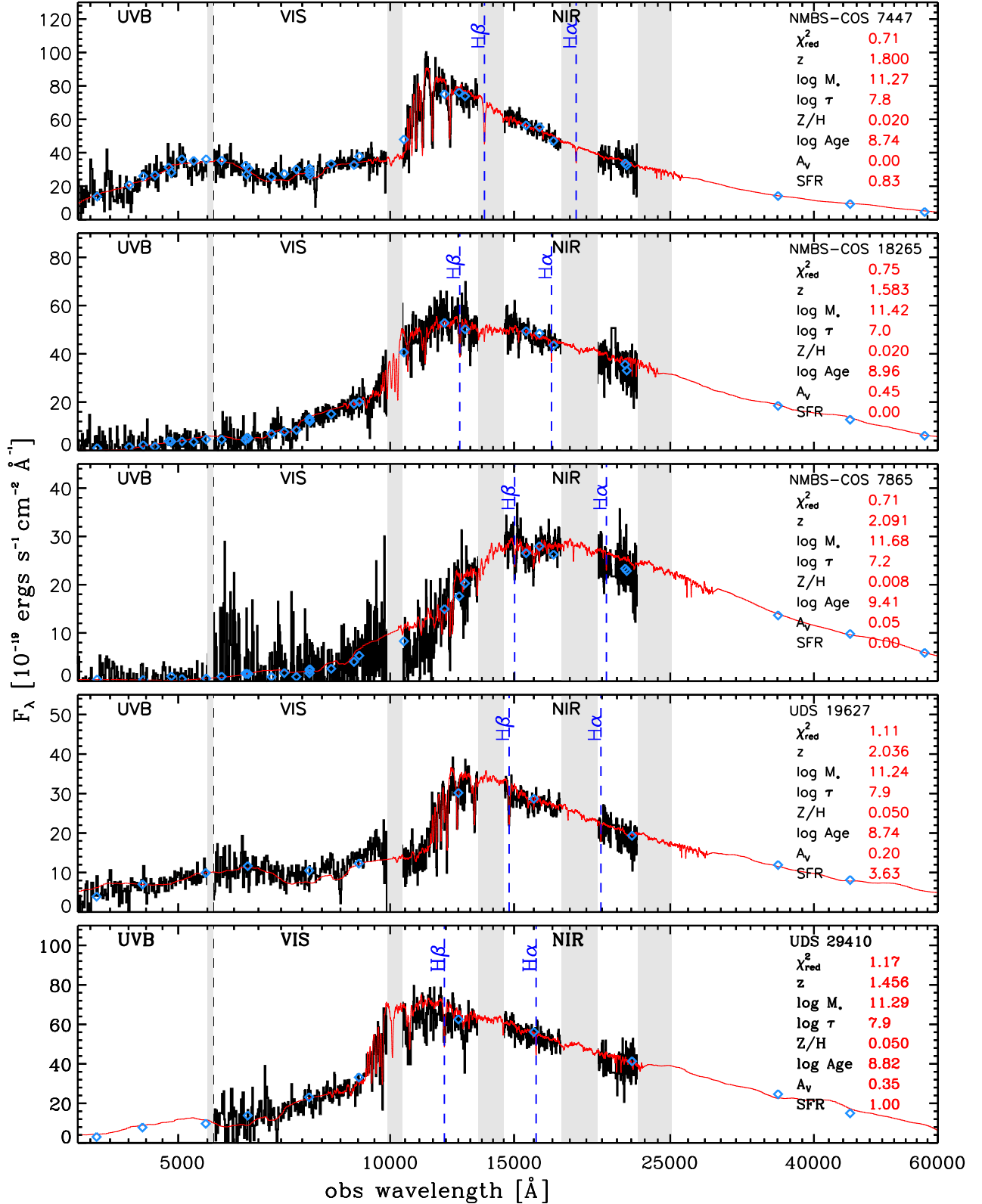}
\caption{UV to NIR X-Shooter spectra in combination with medium- and
broad-band data (blue diamonds). The binned spectra ($\sim10\AA$) are
shown in black, together with the best-fitting BC03 $\tau$-model as
shown in red. Gray areas indicate regions with strong atmospheric
absorption.  The UVB spectrum is missing for UDS-29410, due to an
instrument problem during the observations. The good agreement between
the BC03 models and the spectroscopic data over this large wavelength
range is astonishing. Both NMBS-COS7865 and UDS-19627 show a small
deviation from the best-fitting model around $1\micron$, which is
caused by the absence of good telluric calibrators. From stellar
population synthesis modeling, we find a variety of ages that range
from 0.5-2 Gyr. We find no emission lines, and other signs of star
formation, and with little to no dust (see Section \ref{subsec:sps}).}
\label{fig:spectra_page_1}
\end{figure*}
\begin{figure*}
\epsscale{1.1}
\plotone{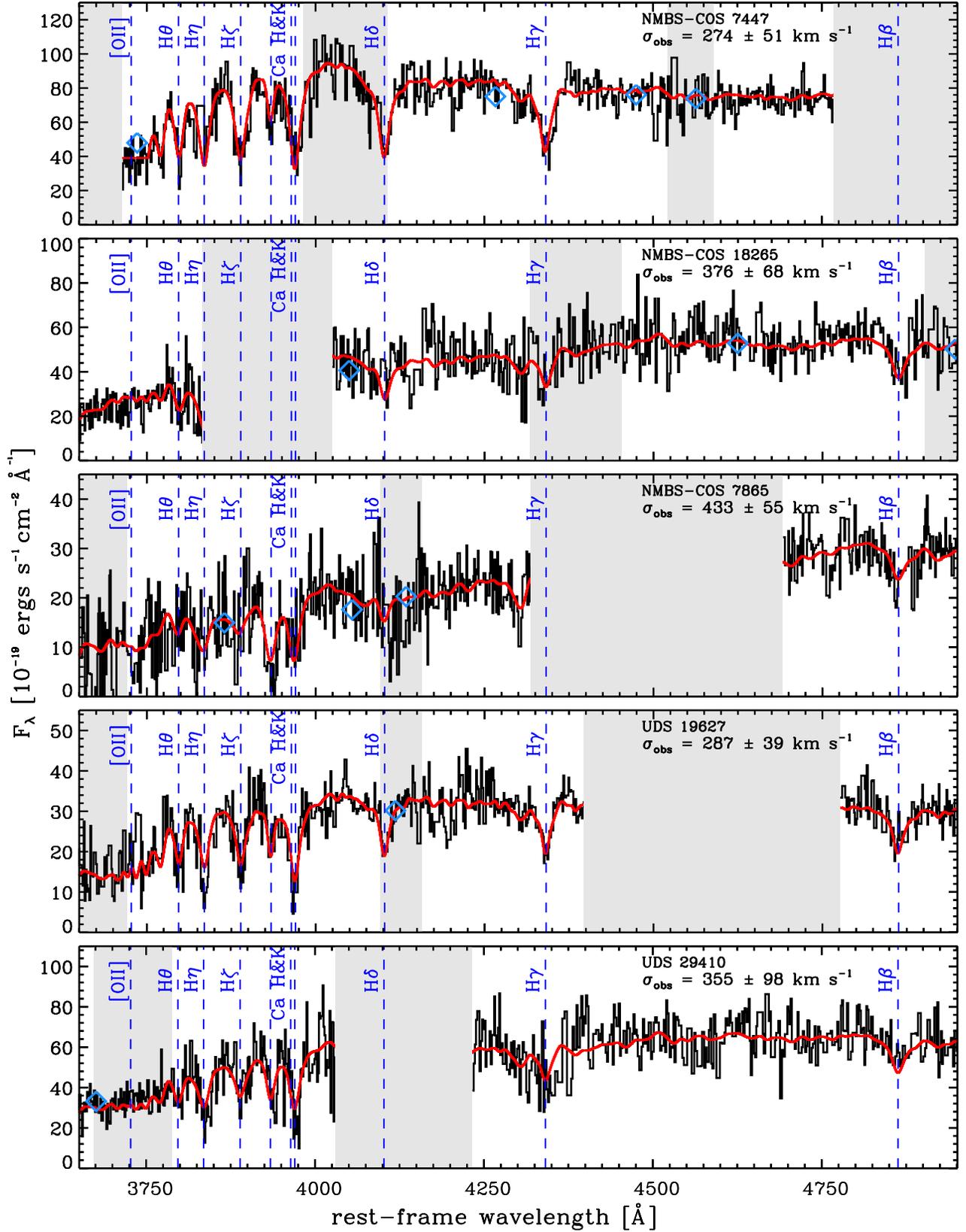}
\caption{Rest-frame optical part of the spectrum focused on the Balmer
break. As in Figure \ref{fig:spectra_page_1}, the X-Shooter spectrum
is shown in black, but this time in higher resolution ($\sim 4\AA$
observed, or $\sim100$ km s$^{-1}$ rest-frame). The most prominent
absorption and emission features are indicated by the blue dashed
lines. The clear detection of absorption lines enables us to measure
stellar velocity dispersions. We use pPXF to fit the best-fitting BC03
$\tau$ model to the spectrum and find velocity dispersions that range
from 275-435 km s$^{-1}$. The convolved best-fit BC03 template is
shown in red.}
\label{fig:spectra_page_2}
\end{figure*}
\begin{figure*}
\epsscale{1.1}
\plotone{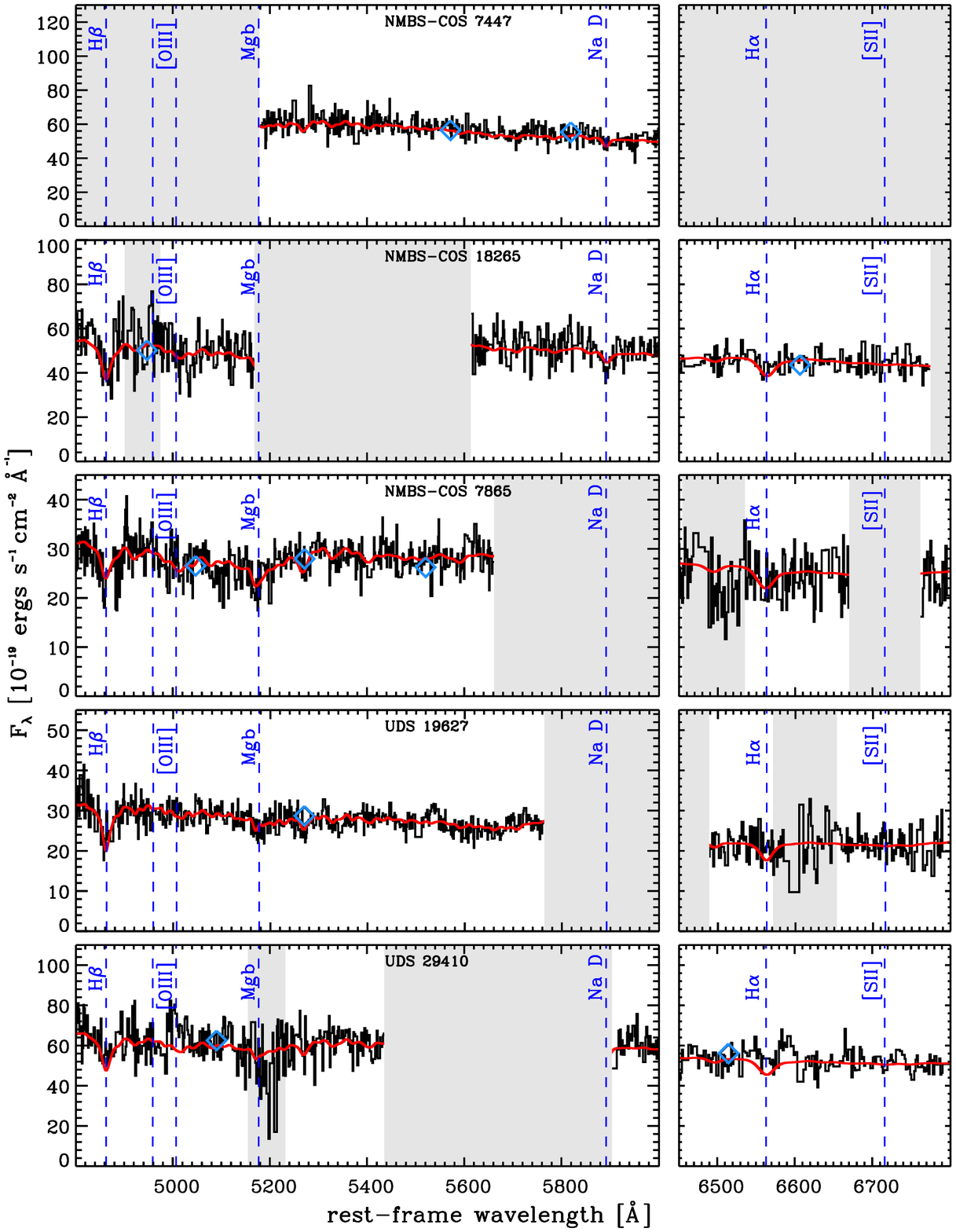}
\caption{Rest-frame optical part of the spectrum focused on $Mgb, Na
D$, and $H\alpha$. As in Figure \ref{fig:spectra_page_2}, the
X-Shooter spectrum is shown in black, with high resolution of $\sim
4\AA$ observed, or $\sim100$ km s$^{-1}$ rest-frame. The most
prominent absorption and emission features are indicated by the blue
dashed lines. The convolved best-fit BC03 template is shown in red.}
\label{fig:spectra_page_3}
\end{figure*}

\section{Data}
\label{sec:data}

\subsection{Target Selection} 
\label{sec:target_selection}
The galaxies in this paper are drawn from the \mbox{NMBS-I}
\citep{whitaker2010} and the UKIDSS-UDS \citep{williams2009}. They
were selected to be bright in the $H$-band, and to have $z >1.4$, in
order to obtain sufficient S/N. The SED
from the broadband and medium-band photometry was required to indicate
that they have quiescent stellar populations, and the rest-frame
optical imaging could not show signs of large disturbance due to
e.g.,mergers. We note that NMBS-COS7447 was presented in
\citet{vandesande2011}, and UDS-19627 was presented in
\citet{toft2012}. All data for both galaxies have been re-analyzed
according to the following procedure for consistency. Our selection
had no priors on mass or size, but could be biased in either one of
these parameters. Full information on the photometric properties of the targets is
listed in Table \ref{tab:phot_sample}.

To investigate possible biases, we compare our targets to a sample of
galaxies with mass $>10^{10.5} M_{\odot}$ at $1.4<z<2.1$ from the
NMBS-I and the UDS. Rest-frame $U-V$ and $V-J$ colors are commonly
used to distinguish between star-forming and quiescent galaxies at
this redshift (e.g.,\citealt{williams2009}). Figure
\ref{fig:selection_effects_sp}$a$ shows the UVJ-diagram for all
galaxies at redshifts between $1.4<z<2.1$ with mass $>10^{10.5}
M_{\odot}$, together with the sample presented here.  The sizes of the
squares indicate the density of galaxies. For our targets, the
rest-frame colors have been measured from the spectra, while for the
full sample rest-frame colors are based on the broadband and medium-band
data. As demonstrated by \citet{williams2009}, non-star-forming
galaxies can be identified using a color selection indicated by the
black lines. Within this selection region, our targets fall in the
area occupied by young, quiescent galaxies \citep{whitaker2012}. The
median specific star formation rates (sSFR), as indicated by the
different colors, are in good agreement with the full high-redshift
sample at the same place in the UVJ diagram. For their mass, however,
NMBS-COS7447 and UDS-19627 have slightly bluer colors as compared to
the full sample (Figure \ref{fig:selection_effects_sp}$b$). At fixed
mass, the targets are among the brightest galaxies, except for
NMBS-I-7865 (\ref{fig:selection_effects_sp}$c$). This may not come as
a surprise as they are among the youngest quiescent galaxies, and thus
have relatively low $M/L$.

\subsection{Imaging} 
\label{subsec:imaging}
Four different imaging data sets are used to measure the surface
brightness profiles of our galaxies, as summarized below. (1) All our
targets in the NMBS-I COSMOS field were observed with HST-WFC3
$H_{160}$ as part of the program HST-GO-12167 (PI: Franx). Each target
was observed for one orbit (2611 sec), using a four point dither
pattern, with half pixel offsets. Reduction of the data was done in a
similar way as to the reduction described in \citet{bouwens2010}, but
without sigma-clipping in order to avoid masking the centers of
stars. The drizzled images have a pixel scale of $0''.06$, with a full
width at half-maximum (FWHM) of the point-spread function (PSF) of
$\sim0''.16$. (2) Our NMBS-I targets are complemented with HST-ACS
$I_{814}$ imaging from COSMOS (v2.0, \citealt{koekemoer2007};
\citealt{massey2010}), which has a $0''.03$ pixel scale and PSF-FWHM
of $\sim0''.11$. (3) For UDS-29410 we make use of the HST-ACS F814W,
HST-WFC3 $J_{125}$ and $H_{160}$ from UDS-CANDELS
(\citealt{grogin2011}; \citealt{koekemoer2011}). These data have the
same properties as the data described in (1) and (2). (4) For
UDS-19627 we use ground based data from UKIDSS-UDS,
(\citealt{lawrence2007}; \citealt{warren2007}) Data Release 8 in the
J, H, and K-band, as no as HST data is available. Imaging in all three
bands were drizzled to a pixel scale $0''.134$, and the FWHM of the PSF
is $0''.7$ in the K-band.

\subsection{Spectroscopic Observations}
Observations were performed with X-Shooter on the VLT UT2
(\citealt{dodorico2006}; \citealt{vernet2011}). X-Shooter is a second
generation instrument on the VLT that consists of three arms: UVB, VIS,
and NIR. The wavelength coverage ranges from 3000 to 24800 $\AA$ in
one single exposure. The galaxies were observed in both visitor and
service mode, and the observations were carried out between 2010 January 
and 2011 March (Programs: Fynbo 084.A-0303(D), Van de Sande
084.A-1082(A), Franx 085.A-0962(A), Toft 086.B-0955(A)). Full
information on the targets and observations is listed in Table
\ref{tab:sample_cat}.  All observations had clear sky conditions and
an average seeing of $0''.$. A $0''.9$ slit was used in the NIR, except
for the 1st hour of UDS-19627 where a $0.''6$ slit was used.  For the
$0''.9$ slit, this resulted in a spectral resolution of 5100 at
1.4$\mu$m. Observing blocks were split into exposures of 10-15
minutes each with an $ABA'B'$ on-source dither pattern. For most
targets, a telluric standard of type B8V-B9V was observed before and
after our primary target, in order to create a telluric absorption
spectrum at the same airmass as the observation of our target.

\subsection{Spectroscopic Reduction}
Data from the three arms of X-Shooter must be analyzed separately and
then combined to cover the full range from the UV to NIR.  In the NIR we
identified bad pixels in the following way. The data were corrected
for dark current, flatfielded, and sky subtracted using the average of
the preceding and subsequent frames.  The ESO pipeline (ver. 1.3.7;
\citealt {goldoni2006}) was used to derive a wavelength solution for
all orders. The orders were then straightened using integer pixel
shifts to retain the pixels affected by cosmic-rays and bad
pixels. Additional sky subtraction was done on the rectified orders,
by subtracting the median in the spatial direction. Cosmic rays and
bad pixels were identified by LA-Cosmic \citep{vandokkum2001}, and a
bad pixel mask was created.

Further 3$\sigma$ clipping was done on the different exposures,
corrected for dithers, to identify any remaining outliers. The bad
pixel masks of different orders were combined into a single file and
then transformed back to the raw frame for each exposure. Masks will
follow the same rectification and wavelength calibration steps as the
science frames.

Next the flatfielded and sky subtracted observations were rectified
and wavelength calibrated, only this time we used interpolation when
rectifying the different orders. Again, additional sky subtraction was
done. Per order, all exposures were combined, with exclusion of bad
pixels and those contaminated with cosmic rays present in the mask
file.

The telluric spectra were reduced in the same way as the science
frames. We constructed a response spectrum from the telluric stars in
combination with a stellar model for a B8V/B9V star from a blackbody
curve and models from \citet{munari2005}. Residuals from Balmer
absorption features in the spectrum of the tellurics were removed by
interpolation. All the orders of the science observations were
corrected for instrumental response and atmospheric absorption by
dividing by the response spectrum.

The different orders were then combined, and in regions of overlap
weighted using the S/N of the galaxy spectrum. A noise spectrum was
created by measuring the noise in the spatial direction below and
above the galaxy. If the regions exceeded an acceptable noise limit,
from contamination by OH lines or due to low atmospheric transmission,
this spatial region was discarded for further use. The two-dimensional (2D) spectra were
visually inspected for emission lines, but none were found.  A 1D
spectrum was extracted by adding all lines (along the wavelength
direction), with flux greater than 0.1 times the flux in the central
row, using optimal weighting.

Absolute flux calibration was performed by scaling the spectrum to the
available photometric data. The scaling was derived for each
individual filter that fully covered the spectrum. For our targets in
NMBS-I we used the following filters: $J, J2, J3, H, H1, H2, and Ks$,
while for the targets in the UDS we only used $J$ and $H$. We then
used an error-weighted average obtained from the broadband magnitudes,
and scaled the whole spectrum using this single value. After scaling,
no color residuals were found, and no further flux corrections were
applied to the spectrum.

\begin{deluxetable*}{c c c c c c c c c c c}[!hb]
\tablecaption{Stellar Population Synthesis Properties}
\tablehead{
\colhead{Catalog} & \colhead{ID} & \colhead{$z_{phot}$} & \colhead{$z_{spec}$} & \colhead{$\log \tau$} & \colhead{$\log Age$} & \colhead{$Z$ } & \colhead{$A_{V}$} & \colhead{$\log M_{*}$} & \colhead{$\log$ SFR} & \colhead{$\log$ sSFR}\\  
& & & & \colhead{(yr)} & \colhead{(yr)} & & \colhead{(mag)} & \colhead{($M_{\odot}$)} & \colhead{($M_{\odot}\mathrm{yr}^{-1}$)} & \colhead{($\mathrm{yr}^{-1}$)} \\ }
\startdata
 NMBS-COS  & 7447  &   1.71$\pm$   0.03  &  1.800  &   7.80  &   8.74  &  0.020  &   0.00  &  11.27  &  -0.08  & -11.35 \\
 NMBS-COS  &18265  &   1.60$\pm$   0.03  &  1.583  &   7.00  &   8.96  &  0.020  &   0.45  &  11.42  & -99.00  & -99.00 \\
 NMBS-COS  & 7865  &   2.02$\pm$   0.05  &  2.091  &   7.20  &   9.41  &  0.008  &   0.05  &  11.68  & -99.00  & -99.00 \\
      UDS  &19627  &   1.94$\pm$   0.06  &  2.036  &   7.90  &   8.74  &  0.050  &   0.20  &  11.24  &   0.56  & -10.68 \\
      UDS  &29410  &   1.44$\pm$   0.02  &  1.456  &   7.90  &   8.82  &  0.050  &   0.35  &  11.29  &   -99.00  & -11.28 \\
\enddata
\tablecomments{Derived stellar population synthesis properties from
FAST. We use stellar templates from \citealt{bruzual2003}, with an
exponentially declining star formation history with timescale $\tau$,
together with a \citet{chabrier2003} IMF, and the \citet{calzetti2000}
reddening law. No errors are provided, as the $68\%$ confidence all
fall within one grid point. The real errors are dominated by
systematic uncertainties.}
\label{tab:stellar_prop}
\end{deluxetable*}

A low resolution spectrum was constructed by binning the 2D spectrum
in wavelength direction. Using a bi-weight mean, 20 good pixels,
i.e.,not affected by skylines or strong atmospheric absorption, were
combined. The 1D spectrum was extracted from this binned 2D spectrum
in a similar fashion as the high resolution spectrum \textit{(see
Figure \ref{fig:spectra_page_1} - \ref{fig:spectra_page_3} )}.

For the UVB and VIS arms we used the ESO reduction pipeline (version
1.3.7, \citealt {goldoni2006}) to correct for the bias, flatfield, and
dark current, and to derive the wavelength solution. The science
frames were also rectified using the pipeline, but thereafter, treated
in exactly the same way as the rectified 2D spectra of the NIR arm, as
described above.

\section{ Structural Properties and Stellar populations}
\label{sec:properties}

\subsection{Stellar Population Properties}
\label{subsec:sps}

We estimate the stellar population properties by fitting the
low-resolution ($\sim10\AA$ in the observed frame) spectrum in the Visual
and NIR in combination with the broadband and medium-band photometry with
SPS models. We exclude the UVB part of the spectrum due to the lower
S/N and the extensive high S/N broadband photometry in this
wavelength region. Stellar templates from \citeauthor{bruzual2003} (\citeyear{bruzual2003}, BC03)
are used, with an exponentially declining star formation history (SFH) with
timescale $\tau$, together with a \citet{chabrier2003} initial mass
function (IMF), and the \citet{calzetti2000} reddening law.

Using the FAST code \citep{kriek2009a}, we fit a full grid in age, dust
content, star formation timescale, and metallicity. We adopt a grid
for $\tau$ between 10 Myr and 1 Gyr in steps of 0.1 dex. The age range
can vary between 0.1 Gyr and 10 Gyr, but the maximum age is
constrained to the age of the universe at that particular redshift. We
note, however, that this constraint has no impact on our results as
the galaxies are young.  Step size in age is set as high as the BC03
templates allow, typically 0.01 dex. Metallicity can vary between
$Z=0.004$ (subsolar), $Z=0.08$, $Z=0.02$(solar), and $Z=0.05$
(supersolar). Furthermore, we allow dust attenuation to range from 0
to 2 mag with step size of 0.05. The redshift used here is from
the best-fitting velocity dispersion (see Section
\ref{subsec:dispersions}).  Results are summarized in Table
\ref{tab:stellar_prop}.

Due to our discrete grid and the high quality data, and also because
metallicity and age are limited by the BC03 models, our $68\%$
confidence levels are all within one grid point.  Our formal errors
are therefore mostly zero, and not shown in Table
\ref{tab:stellar_prop}. This does not reflect the true uncertainties,
which are dominated by the choice of SPS models, IMF, SFH, and extinction
law (see e.g.,\citealt{conroy2009}; \citealt{muzzin2009}).

The low sSFR confirms the quiescent nature of the galaxies in our
sample, and they match well with the sSFR of the general population in
the same region of the UVJ diagram (Figure
\ref{fig:selection_effects_sp}$a$). We find a range of metallicities,
with the oldest galaxy having the lowest metallicity. However, due to
the strong degeneracy between age and metallicity, we do not believe
this result to be significant. Overall, the dust content in our
galaxies is low.

We find very similar results for NMBS-C7447 as compared to
\citet{vandesande2011}, and the small differences can be explained by
the newer reduction. For UDS-19627 we find a slightly lower mass as
compared to \citet{toft2012}, which is likely due to the lower dust
fraction that we find, i.e $A_v=0.2$ vs. $A_v=0.77$ from
\citet{toft2012}.

The galaxies in our sample are not detected at 24 $\mu$m, leading to a
3-$\sigma$ upper-limit of 18 \ $\mu$Jy for the galaxies in
NMBS-COSMOS, and 30\ $\mu$Jy for UDS-19627 (see
\citealt{whitaker2012}; \citealt{toft2012}). UDS-29410 has a strong
detection at 24 $\mu$m of $232 \pm 15 \mu Jy$.  From these upper
limits we calculate the dust-enshrouded SFRs that are listed in Table
\ref{tab:sample_cat}. We find a high SFR for UDS-29410, but we find no
other signs for this high SFR. That is, we find no emission lines and
the best fitting SPS model indicates a low SFR. Therefore, we think
that the strong 24 $\mu$m detection is likely due to an obscured AGN.

\begin{deluxetable*}{c c c c c c c c c c c c c c c c c c c}[!th]
\tabletypesize{\scriptsize}
\tablewidth{0pt}
\tablecaption{Compilation of Masses and Structural Parameters for High-Redshift Galaxies}
\tablehead{
\colhead{Reference$^\mathrm{a}$} & \colhead{ID} & \colhead{$z_{spec}$} & \colhead{$r_e$}  & \colhead{$n_{sersic}$}   & \colhead{${b/a}$}   & \colhead{$\sigma_{e}$} & \colhead{$\sigma_{e} / \sigma_{ap}$} & \colhead{$\beta(n)$} & \colhead{$\log M_{dyn}$} & \colhead{$\log M_{*,corr}$} & \colhead{Filter}}\\  
\startdata
0  &  7447  & 1.800  &  1.75  $\pm$  0.21  &  5.27  $\pm$  0.23  &  0.71  $\pm$0.02  & 287$^{+ 55}_{- 52}$ &  1.048  &  5.16  & 11.24$^{+0.13}_{-0.12}$ & 11.22  & $H_{\rm F160W}$  \\
0  & 18265  & 1.583  &  0.97  $\pm$  0.12  &  2.97  $\pm$  0.06  &  0.26  $\pm$0.02  & 400$^{+ 78}_{- 66}$ &  1.065  &  6.61  & 11.38$^{+0.13}_{-0.11}$ & 11.32  & $H_{\rm F160W}$   \\
0  &  7865  & 2.091  &  2.65  $\pm$  0.33  &  4.82  $\pm$  0.15  &  0.83  $\pm$0.02  & 446$^{+ 54}_{- 59}$ &  1.031  &  5.42  & 11.82$^{+0.09}_{-0.09}$ & 11.64  & $H_{\rm F160W}$   \\
0  & 19627  & 2.036  &  1.32  $\pm$  0.17  &  3.61  $\pm$  0.73  &  0.48  $\pm$0.06  & 304$^{+ 43}_{- 39}$ &  1.059  &  6.18  & 11.24$^{+0.10}_{-0.09}$ & 11.20 & $K$   \\
0  & 29410  & 1.456  &  1.83  $\pm$  0.23  &  2.59  $\pm$  0.03  &  0.54  $\pm$0.02  & 371$^{+114}_{- 90}$ &  1.045  &  6.88  & 11.61$^{+0.19}_{-0.15}$ & 11.24  & $H_{\rm F160W}$   \\
\nodata & \nodata & \nodata & \nodata & \nodata & \nodata & \nodata & \nodata & \nodata & \nodata & \nodata \nodata \\
\enddata
\tablecomments{This Table will be published in its entirety in the
electronic edition of ApJ, and can also be downloaded from
\url{http://www.strw.leidenuniv.nl/~sande/data/}.  A portion is shown
here for guidance regarding its form and content.
Spectroscopic redshifts $z_{spec}$ are obtained from the velocity
dispersion fit as described in Section
\ref{subsec:dispersions}. Structural parameters $r_e$, $n_{sersic}$,
and $b/a$ are derived using GALFIT on available imaging, as described
in Section \ref{subsec:sizes}. $\sigma_e$ is the velocity dispersion
within a circular aperture of size $r_e$ from Section
\ref{subsec:dispersions}, and $\sigma_{e} / \sigma_{ap}$ is the
aperture correction we apply to the observed velocity dispersion as
described in Appendix \ref{sec:app_aper_corr}. From Equation
\ref{eq:kn} we calculated $\beta (n)$, and dynamical masses are
derived using Equation \ref{eq:mdyn}. Stellar masses as given here are
corrected to account for the difference between the catalog magnitude
and our measured magnitude. The filter in which the structural
parameters are measured is given in the last column.
 }
\tablenotetext{a}{References: 0) this work 
1) \citet{bezanson2012}; 
2) \citet{vandokkum2009a};
3) \citet{onodera2012};
4) \citet{cappellari2009};
5) \citet{newman2010};
6) \citet{vanderwel2008} and \citet{blakeslee2006};
7) \citet{toft2012}.} 
\label{tab:sample_results}
\end{deluxetable*}


\subsection{Velocity Dispersions}
\label{subsec:dispersions} 

The clear detection of absorption lines in our spectra, together with
the medium resolution of X-Shooter, enable the measurement of accurate
stellar velocity dispersions. We use the unbinned spectra in
combination with the Penalized Pixel-Fitting (pPXF) method by
\citet{cappellari2004} and our best-fitting BC03 models as templates.
Spectra were resampled onto a logarithmic wavelength scale without
using interpolation, but with masking of the bad pixels. The effect of
template mismatch was reduced by simultaneously fitting the template
with a $\sim$17-order Legendre Polynomial. Our results depend only
slightly on the choice of the order of the polynomial (Appendix
\ref{sec:app_disp_test}). Together with the measured velocity
dispersion, the fit also gives us the line-of-sight velocity, and thus
$z_{\rm spec}$.

We also look at dependence of the velocity dispersion on the template
choice.  In particular for the younger galaxies in our sample that
show a clear signature of A-type stars, we find a dependence of the
measured velocity dispersion as a function of template age. A more
stable fit is obtained when restricting the wavelength range to $4020
\AA < \lambda <7000 \AA$, which excludes the Balmer break region (see
also Appendix \ref{sec:app_disp_test}).
 
The errors on the velocity dispersion were determined in the following
way. We subtracted the best-fit model from the spectrum. Residuals are
shuffled in wavelength space and added to the best-fit template. We
then determined the velocity dispersion of 500 of these simulated
spectra. Our quoted error is the standard deviation of the resulting
distribution of the measured velocity dispersions. When we include the
Balmer break region in the fit, the formal random error decreases, but
the derived dispersion becomes more dependent on the chosen stellar
template. In total we have three high-quality measurements, and two
with medium quality. We note that if we exclude the two galaxies with
medium-quality measurements from our sample, our main conclusions
would not change.

The velocity dispersion found here for NMBS-C7447 agrees well with the
results from \citet{vandesande2011}. For UDS-19627 we find a slightly
lower value as compared to \citet{toft2012}.  However, they use a
different method for constructing the template for the velocity
dispersion fit. When we fit the spectrum of UDS-19627 in the same way
as was described in \citet{toft2012}, we find a similar answer as
theirs.

All dispersions are corrected for the instrumental resolution ($\sigma$=23
km s$^{-1}$) and the spectral resolution of the templates ($\sigma$=85
km s$^{-1}$). Furthermore, we apply an aperture correction to our
measurements as if they were observed within a circular aperture
radius of $r_e$. In addition to the traditional correction for the
radial dependence of velocity dispersion
(e.g.,\citealt{cappellari2006}), we account for the effects of the
non-circular aperture, seeing and optimal extraction of the 1-D
spectrum. The aperture corrections are small with a median of 4.8$\%$
(See Appendix \ref{sec:app_aper_corr}).  The final dispersions and
corresponding uncertainties are given in Table
\ref{tab:sample_results}.

\subsection{Surface Brightness Profiles}
\label{subsec:sizes} 
Radial profiles are measured for all galaxies on all available imaging
as described in Section \ref{subsec:imaging}. Galaxies are fitted by
2D S\'ersic radial surface brightness profiles
\citep{sersic1968}, using GALFIT (ver. 3.0.2;
\citealt{peng2010}). Relatively large cutouts of $25''\times 25''$
were provided to GALFIT to ensure an accurate measurement of the
background, which was a free parameter in the fit.  All neighboring
sources were masked using a segmentation map obtained with SExtractor
\citep{bertin1996}. In the case of UDS-19627, the close neighbor was
fitted simultaneously. Bright unsaturated field stars were used for
the PSF convolution. All parameters, including the sky, were left free
for GALFIT to determine.

Even though galaxies at low redshift are well-fitted by single
S\'ersic profiles (e.g.,\citealt{kormendy2009}), this does not
necessarily have to be true for galaxies at $z\sim2$. Therefore, we
correct for missing flux using the method described in
\citet{szomoru2010}. We find very small deviation in residual-corrected 
effective radii, with a median absolute deviation of
3.4$\%$. Color images and measured profiles are shown in Figure
\ref{fig:sizes}.

We repeated the measurements using a variety of PSF stars
($N\sim25$). We find an absolute median deviation in the half-light
radius of $\sim 3\%$ for HST-WFC3, $\sim 3.5\%$ for HST-ACS, and $\sim
10\%$ for the ground-based UDS-UKIDSS data, due to variations in the
PSF. The largest source of uncertainty in the measured profiles is,
however, caused by the error in the sky background estimate. Even
though these galaxies are among the brightest at this redshift, using
the wrong sky value can result in large errors for both $r_e$ and
$n$. We determine the error in the sky background estimate by
measuring the variations of the residual flux in the profile between 5
and 15 arcsec. For sizes derived from HST-ACS, the absolute median
deviation in the effective radius due to the uncertainty in background
is $\sim 13\%$, and for HST-WFC3 $\sim 12\%$. Due to the deeper
ground-based UDS-UKIDSS data, the uncertainty for UDS-19627 due to the
sky is $\sim 8\%$. All of our results are summarized in Table
\ref{tab:sample_results}.

We note that we find a smaller size and larger $n$ for UDS-19627 as
compared to \citet{toft2012}, which cannot be explained within the
quoted errors. We have compared our results with the size measurements
from \citet{williams2009} and R. J. Williams (2012, private communication), 
who also use UDS-UKIDSS data for measuring structural
parameters. They too find a smaller size in the $K$ band of $r_e =
1.63\,$kpc, with a similar axis ratio of $q = 0.53$, while keeping the
S\'ersic index fixed to $n=4$. Furthermore, we compare the size of
UDS-29410 obtained from the ground-based UDS-UKIDSS data, to the size
from HST-WFC3 to test how reliable the ground-based data are for
measuring structural parameters. From the ground-based UDS $H$ band we
find $r_e = 1.97 \pm 0.11$ kpc, and $n=2.47 \pm 0.22$ for UDS-29410
which is consistent with the measurements using the HST-WFC3 data
within our $1-\sigma$ errors. From these two independent results, we
are confident that our size measurement for UDS-19627 is correct.

In what follows, we will use the mean
effective radius and S\'ersic $n$ from the band which is closest to
rest-frame optical $r'$. The effective radii reported here are circularized, $r_e = \sqrt{ab}$.

\subsection{Dynamical Masses}
\label{subsec:mdyn} 

Combining the size and velocity dispersion measurements we are now
able to estimate dynamical masses using the following expression:
\begin{equation} 
M_{\rm dyn}=\frac{\beta(n)~ \sigma_{e}^2~r_e }{G}.
\label{eq:mdyn}
\end{equation}
Here $\beta(n)$ is an analytic expression as a function of the
S\'ersic index, as described by \citet{cappellari2006}:
\begin{equation} 
\beta(n) = 8.87 - 0.831n + 0.0241n^2.
\label{eq:kn}
\end{equation}
This is computed from theoretical predictions for $\beta$ from
spherical isotropic models described by the S\'ersic profile, for
different values of $n$, and integrated to one $r_e$
(cf. \citealt{bertin2002}). The use of a S\'ersic-dependent virial
constant $\beta(n)$ gives a better correspondence between $M_{dyn}$
and $M_{*}$ for galaxies in the SDSS \citep{taylor2010b}. This does
require however, that the total stellar masses are also derived using
the luminosity of the derived S\'ersic profile.  Thus we correct our
total stellar mass, as derived from the total magnitude as given in
the catalogs (measured with Sextractor), to the total magnitude from
the S\'ersic fit.  We note that the values for $\beta$ that we find
are all close to 5, a value often used in the literature
(e.g.,\citealt{cappellari2006}). Our dynamical masses and corrected
stellar masses are given in Table \ref{tab:sample_results}.

\section{Compilation of Kinematic Studies}
\label{sec:compilation}

In order to study the structural evolution of quiescent galaxies,
we combine the results from different kinematic studies at
various redshifts. Where possible, we apply similar corrections as
described above.

\begin{figure}
\epsscale{1.1}
\plotone{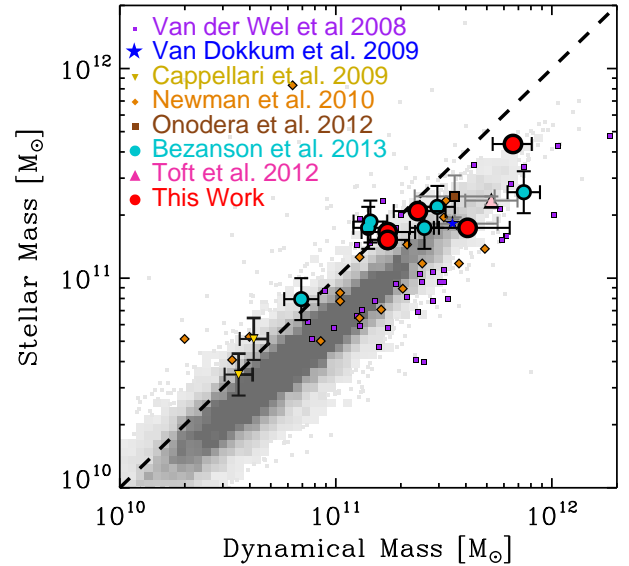}
\caption{Comparison of the stellar mass vs. the
dynamical mass. Gray squares are non-star-forming galaxies from
the SDSS. Different symbols are from a compilation of high-redshift
galaxies as described in Section \ref{sec:sample_highz}. The dashed
line is for equal dynamical and stellar mass. Low-redshift galaxies
are all below the line, as is expected given the contribution of dark
matter. All our high-redshift galaxies have dynamical masses that are
close to the stellar mass. This suggests that the stellar
mass measurements at high-redshift are robust for passive galaxies.}
\label{fig:mdyn_mstar}
\end{figure}

\begin{figure*}
\epsscale{1.1}
\plotone{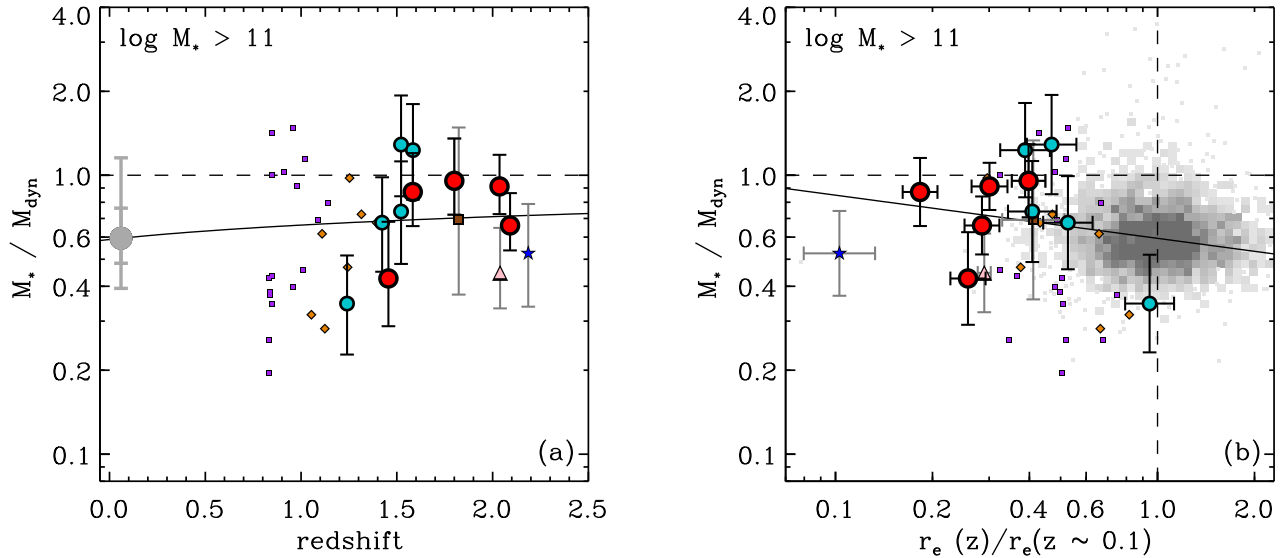}
\caption{(a) Stellar mass divided by the dynamical
mass vs. redshift. Galaxies below the line have dynamical mass
greater than the stellar mass, above the line is the non-physical
regime. For SDSS galaxies with stellar mass $>10^{11} M_{\odot}$, we
find that they have a median $M_*/M_{dyn}$ of 0.59. Up to redshift
$z=1.5$ we find a similar slightly lower median value ($\sim0.5$), but
it rapidly increases at $z>1.5$ with median a of $M_{*}/M_{dyn} =
0.9$. The solid line is the best-fit $M_{*}/M_{dyn}\propto
(1+z)^{0.17\pm0.11}$. We caution that this result might be biased due
to the selection effects as explained in Section \ref{sec:properties},
and relatively large measurement errors. (b) Stellar
mass divided by the dynamical mass vs. the evolution in the
effective radius at fixed dynamical mass. Galaxies that have small
effective radii at fixed dynamical mass also show higher ratios of
$M_{*} / M_{dyn}$, although there is a significant scatter. The solid
line is the best-fit $M_{*} / M_{dyn} \propto \left(r_e(z) \, / \,
r_e(z\sim0.1)\, \right)^{0.16\pm0.10}$.}
\label{fig:redshift_delta_mass}
\end{figure*}


\subsection{Low-Redshift Sample}
\label{sec:sample_sdss}
At low redshift we select galaxies from the SDSS DR7. Stellar masses are based on
MPA\footnote{http://www.mpa-garching.mpg.de/SDSS/DR7/} fits to the
photometry following the method of \citet{kauffmann2003}, and
\citet{salim2007}. Star formation rates (SFRs) are based on
\citet{brinchmann2004}. Structural parameters are from the NYU
Value-Added Galaxy Catalog (NYU-VAGC, \citealt{blanton2005}). For all
galaxies, velocity dispersions were aperture corrected as described in
Section \ref{subsec:dispersions}, and stellar masses are calculated
with a \citet{chabrier2003} IMF. We furthermore correct the stellar
masses using the total magnitude from the best S\'ersic fit. All
dynamical masses were derived using Equation \ref{eq:mdyn}. For making
an accurate comparison between low- and high-redshift galaxies, we
only select non-star-forming galaxies, i.e sSFR $ < 0.3/t_H$ (see
\citealt{williams2009}), where $t_H$ is the age of the universe at the
given redshift.

\subsection{Intermediate- and High-redshift Sample}
\label{sec:sample_highz}
Our high redshift sample consists of a collection of both optical and
NIR spectroscopic studies of individual
galaxies. \citet{vanderwel2008} present a sample of quiescent galaxies
at $z\sim1$, which itself is a compilation of three studies in the
following fields: Chandra Deep Field South (CDF-S;
\citealt{vanderwel2004}; \citeyear{vanderwel2005}), the Hubble Deep
Field North (HDF-N; \citealt{treu2005a}; \citeyear{treu2005b}), and
cluster galaxies in MS 1054-0321 at $z=0.831$ \citep{wuyts2004}. We
derive stellar masses for this sample by running the stellar
population code FAST on available catalogs, i.e., FIREWORKS
\citep{wuyts2008} for the CDS-S, \citet{skelton2012} for the HDF-N,
and FIRES \citep{forsterschreiber2006} for MS 1054-0321. For CDF-S and
HDF-N the stellar masses are corrected using the total magnitude from
the best $n=4$ fit to be consistent with the structural parameters
from \citet{vanderwel2008}. For MS 1054-0321, we use structural
parameters and stellar mass corrections based on the results by
\citet{blakeslee2006}, who fit S\'ersic profiles with $n$ as a free
parameter. We note that \citet{martinez2011} also study a sample of
four $z\sim1$ galaxies, but find dynamical masses that are
significantly lower than their stellar masses, in contrast to the
result by \citet{vanderwel2008}.

Other high-redshift results included here are from \citet{newman2010}
and \citet{bezanson2012}, who use the upgraded red-arm of LRIS on Keck
to obtain UV rest-frame spectra of galaxies at $z \sim1.3$ and $z \sim
1.5$ respectively. Velocity dispersions for two galaxies at $z=1.41$
are presented by \citet{cappellari2009}, and have been observed with
VLT-FORS2 (see also \citealt{cenarro2009}). Using NIR spectrographs,
\citeauthor{onodera2012} (\citeyear{onodera2012}, Subaru-MOIRCS) and
\citeauthor{vandokkum2009a} (\citeyear{vandokkum2009a}, GNIRS)
obtained velocity dispersions for two galaxies at $z=1.82$ and
$z=2.186$. Similar to the current study, \citet{toft2012} study
UDS-19627 using VLT X-Shooter. Dynamical masses were derived using to
Equation \ref{eq:mdyn}. Note that for the studies of
\citet{cappellari2009}, \citet{onodera2012}, \citet{vandokkum2009a},
and \citet{toft2012} no stellar mass corrections were applied due to
the absence of the necessary information. All structural and kinematic
properties of our high-redshift sample are listed in Table
\ref{tab:sample_results}.

\section{Are stellar masses reliable?}
\label{sec:mass_comparison}

The main goal of this paper is to see whether the stellar masses at
$z\sim2$ are reliable. Here we compare our stellar masses, as derived
from the spectra and photometry, to our dynamical masses, which are
derived from effective radii and stellar velocity dispersions (Figure
\ref{fig:mdyn_mstar}). Gray squares represent the density of
non-star-forming, low-redshift galaxies from the SDSS as described in
Section \ref{sec:sample_sdss}. Other symbols are the high-redshift
studies as described in Section \ref{sec:sample_highz}. The one-to-one
relation for $M_{dyn}$ and $M_*$ is indicated by the dashed line. Note
that the region above the line is nonphysical with stellar masses
being higher than the dynamical mass.

Most $z>1.5$ galaxies in this sample are very massive, in the range
$11.2 <$ log M$_{dyn}/$M$_{\odot} < 11.8$.  At all redshifts, stellar
and dynamical masses are tightly correlated and dynamical mass, which
includes baryonic and dark matter, is on average higher than stellar
mass. Thus, we infer that the stellar masses of our galaxies are
broadly correct, and that the apparent size evolution of massive
galaxies in photometric studies cannot be explained by errors in the
photometric masses (see also \citealt{vanderwel2008}).

Figure \ref{fig:redshift_delta_mass}a shows the ratio of the stellar
and dynamical mass as function of redshift for all galaxies with
stellar mass $>10^{11} M_{\odot}$.  We see that the average ratio at
low-redshift for massive galaxies is a factor of 0.59 with a scatter
of 0.12 dex. We note that For MS 1054-0321, the ratio of the stellar
to dynamical mass are slightly higher as compared to low redshift
galaxies. Up to redshift $z\sim1.5$ we find a similar value
($\sim0.5$) with similar scatter, but at higher redshift, the ratio
seems to decline. For galaxies at $z>1.5$ we find a median ratio of
$M_{*} / M_{dyn} = 0.9$. We quantify the evolution in this ratio by
fitting the relation:
\begin{equation} M_{*} / M_{dyn} \propto (1+z)^{\alpha}.
\label{eq:delta_mass_redshift}
\end{equation}
We use a linear least-squares fit in log-log space using the function
$MPFIT$ \citep{markwardt2009}, which takes the errors on each
individual data point into account. We find a best-fitting value of $
\alpha = 0.17 \pm 0.11$, which is shown as the solid black line in
Figure \ref{fig:redshift_delta_mass}a. The uncertainty is derived from
1000 bootstrap simulations, where we draw data points randomly from
the sample. The quoted error is the standard deviation from the
resulting distribution of points. Even though the fit is statistically
significant at the $1- \sigma$ level, due to the relatively large
measurements errors as compared to low redshift, and the possible
selection bias of the high-redshift samples, we are cautious to draw
any strong conclusions from this result.


\begin{figure*}
\epsscale{1.15}
\plotone{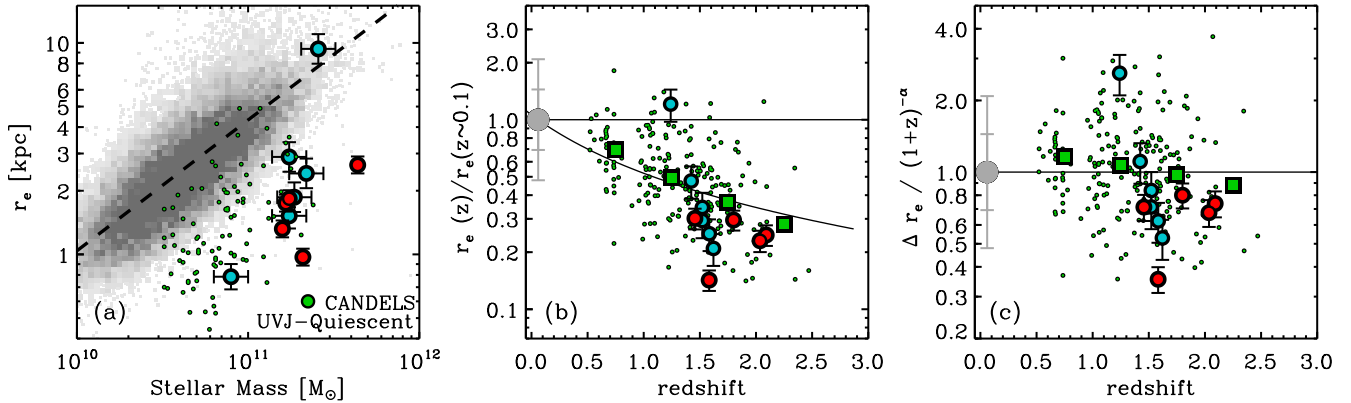}
\caption{Structural comparison of our spectroscopic sample to the full
quiescent galaxy population. (a) Effective radii
vs. mass for low- and high-redshift galaxies. Gray squares are
non-star-forming $z\sim0$ galaxies from the SDSS, with the dashed line
the best-fit to Equation \ref{eq:mstarre}. High-redshift quiescent
galaxies ($1.4 < z< 2.1$) from CANDELS (\citealt{grogin2011};
\citealt{koekemoer2011}) are shown as small green circles, together
with our spectroscopic sample shown as red circles, and the sample by
\citet{bezanson2012} shown as cyan circles. At fixed mass, our
high-redshift galaxies have smaller effective radii, similar to what
has been found by many studies. (b) Evolution in
effective radius at fixed stellar mass, thus corrected for the $M_* -
r_e$ relation in panel (a). Big green squares are the median
effective radii in bins of redshift for the CANDELS data. The solid
line is the best-fit $r_e \propto (1+z)^{-1.02\pm0.05}$. At similar
redshift, we find that our sample and that of \citet{bezanson2012} are
mostly below this fit, indicating that our samples our biased towards
smaller effective radii. (c) Similar to the
panel (b), but now divided by $(1+z)^{-1.02}$ for a better comparison of
our spectroscopic sample to the CANDELS data. When comparing the
median of our sample to the binned median of other quiescent galaxies
at similar redshift, we find smaller effective radii by a factor
$\sim1.28$. This might be explained by our selection which is based on
aperture magnitude, which tends to be biased towards smaller galaxies.
}
\label{fig:selection_effects_mre}
\end{figure*}

It is tempting to speculate that the evolution in $M_*/M_{dyn}$ might
have been caused by a decrease in the dark matter fraction as a
function of redshift.  For galaxies growing in size over time, the
dark matter fraction within $r_e$ will also increase. As the dark
matter profile is less steep than the stellar mass profile, the dark
matter to stellar mass fraction increases with radius, in a similar
fashion as shown here (e.g.,\citealt{hopkins2009a}). If so, this could
also indicate that the IMF at high-redshift is very similar to the IMF
at low-redshift.

Figure \ref{fig:redshift_delta_mass}b shows $M_{*} / M_{dyn}$ vs.
the evolution of the effective radius at fixed dynamical mass (see
Section \ref{sec:size_evolution} and Figure
\ref{fig:redshift_size}). Although there is significant scatter, we do
find the galaxies with high $M_{*} / M_{dyn}$ also tend to have
smallest size at fixed dynamical mass. Galaxies that are closest to
the present-day mass-size relation (dashed vertical line) show lower
ratios of stellar to dynamical mass.

We test this claim by using the the following equation:
\begin{equation} 
M_{*} / M_{dyn}  \propto \left(r_e(z) \, / \, r_e(z\sim0.1)\, \right)^{\alpha}.
\label{eq:delta_mass_re}
\end{equation}
We find $\alpha = -0.16 \pm 0.10$, where the error is determined in a
similar way as described for Equation \ref{eq:delta_mass_redshift} using the bootstrap method
Furthermore, we use the Spearman's rank test on the intermediate- and
high-redshift data. This confirms that there is an anti-correlation
with a probability of 96$\%$. The best-fitting Spearman's rank
correlation coefficient is $-0.28 \pm 0.08$. Even though we find a
weak anti-correlation, this agrees with the idea that the decreasing
ratio of $M_{*} /M_{dyn}$ with time might be correlated to the size
growth of massive galaxies.

\section{Structural evolution of quiescent galaxies}
\label{sec:evolution}

In this section, we will re-examine the structural evolution
of massive quiescent galaxies but now using dynamical measurements.

\subsection{Bias towards Compact Galaxies}
\label{sec:bias}

As noted in Section \ref{sec:target_selection}, this sample is biased
towards young quiescent galaxies. Therefore, we will first investigate
whether our sample and that of \citet{bezanson2012} are biased in size
as compared to other high-redshift galaxies. We gathered structural
properties of galaxies from two studies that use CANDELS data in the
UDS and GOODS-South fields (\citealt{patel2012};
\citealt{szomoru2012}). We compare to a subsample of these galaxies
that are determined to be quiescent from their rest-frame U-V and V-J
colors (see e.g.,Figure \ref{fig:selection_effects_sp}a). When
comparing the effective radii vs. the stellar mass in Figure
\ref{fig:selection_effects_mre}$a$, we find that our galaxies (red
circles) and those of \citeauthor{bezanson2012} (\citeyear{bezanson2012}, cyan circles) are in
general more compact as compared to the high-redshift CANDELS galaxies
(small green circles).

For low-redshift galaxies we parametrize the mass-size relation by:
\begin{equation} 
r_e=r_c \left( \frac{M_{\rm *}}{10^{11}M_{\odot}}\right)^b
 \label{eq:mstarre}
\end{equation}
(\citealt{shen2003}; \citealt{vanderwel2008}). Using a linear
least-squares fit in log-log space, we find best fitting values of
$r_c=4.32~$kpc and $b=0.62$. This is slightly different from the fit
by \citet{shen2003} who find $r_c=4.16~$kpc and $b=0.56$. The
difference may be explained by different selection criteria, and their
use of an older release version of SDSS. Figure
\ref{fig:selection_effects_mre}$b$ shows the evolution in effective
radius, by comparing galaxies with similar mass at different
redshifts. Using both the SDSS and the CANDELS data, we examine the
amount of evolution in size by fitting the following relation:
\begin{equation} 
r_e \propto (1+z)^{\alpha}, 
\label{eq:delta_re_redshift}
\end{equation}
We find $\alpha = -1.02 \pm 0.05$ (linear fit in log-log space). Our
spectroscopic targets and those of \citet{bezanson2012} are mostly
below this best-fit relation, being smaller by a factor of $\sim1.28$
as compared to median in redshift bins (big green squares). This is
especially clear from Figure \ref{fig:selection_effects_mre}$c$, where
we correct for the evolution in size.

This bias might be explained by the method our targets are
selected. As our selection is based on the magnitude within a fixed
aperture of $1''.5$, instead of the total magnitude, we create a bias
towards compact galaxies. For galaxies with similar total magnitudes,
the smaller galaxies will be brighter within a photometric aperture,
and thus make it into our sample.  In what follows, we correct for
this bias by increasing our sizes and those of \citet{bezanson2012} by
a factor of $1.28$, and decreasing the velocity dispersion by a factor
of $\sqrt{1.28}$.
 

\begin{figure*} \epsscale{1.15} \plotone{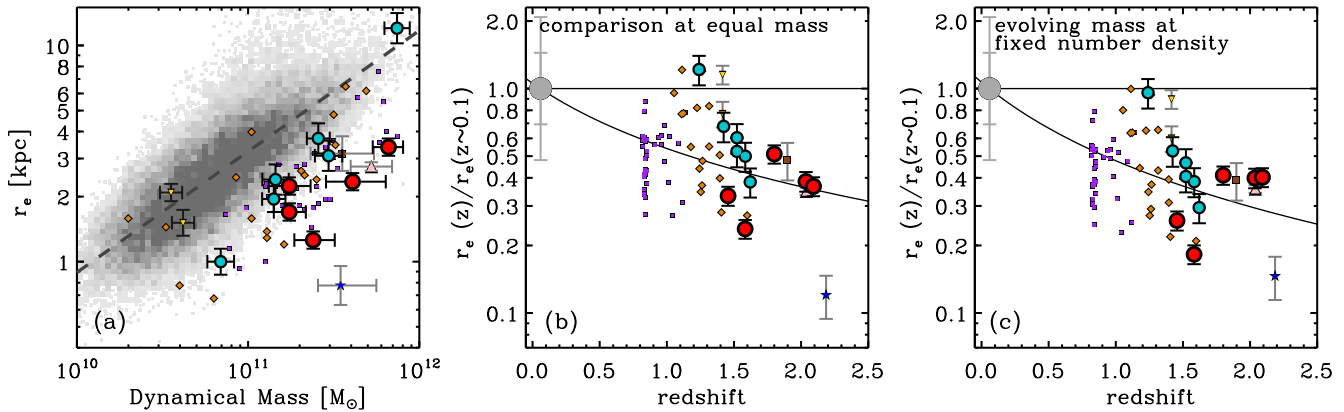}
\caption{Redshift evolution of the effective radius of passive
galaxies. (a) Size vs. dynamical mass. The dashed line
is the best-fit derived using Equation \ref{eq:mdre} for
non-star-forming SDSS galaxies. At fixed dynamical mass, we find that
$z\sim2$ galaxies (red circles) are smaller by a factor $\sim3$
compared to low-redshift galaxies. (b) Evolution of
the effective radius at fixed dynamical mass vs. redshift. The
solid line is the best-fit $r_e \propto (1+z)^{-0.97\pm0.1}$, similar
to what has been found by other stellar kinematic studies at
high-redshift. Similar to Figure \ref{fig:selection_effects_mre}, our
sample is below the best-fit line to the entire high-redshift
sample. (c) Evolution of the effective radius using
an evolving mass function at constant number density. We now compare
galaxies at high-redshift, to more massive galaxies at low-redshift,
assuming the mass evolves as $\Delta \log M/M_{\odot} \sim
0.15z$. This time, we find an even stronger evolution, with $r_e
\propto (1+z)^{-1.16\pm0.1}$.  }
\label{fig:redshift_size}
\end{figure*}

\subsection{Evolution in Size}
\label{sec:size_evolution}

In Figure \ref{fig:redshift_size}a we plot effective radius vs.
dynamical mass. Symbols are the same as in Figure
\ref{fig:redshift_delta_mass}. For the low-redshift galaxies we
parametrize the mass-size relation according to the following equation
\begin{equation} 
r_e=r_c \left( \frac{M_{\rm dyn}}{10^{11}M_{\odot}}\right)^b,
 \label{eq:mdre}
\end{equation}
and find $r_c=3.23~$kpc and $b=0.56$ (dashed line). This is in good
agreement with $b=0.56$ and $r_c=3.26~$kpc as found by
\citet{vanderwel2008}. At fixed dynamical mass, we see that all our
galaxies have smaller effective radii as compared to
low-redshift. This finding is further illustrated in Figure
\ref{fig:redshift_size}b, where we compare the effective radii at
fixed dynamical mass to the mass-size relation at $z\sim0$. The solid
line is the best-fit as described by equation
\ref{eq:delta_re_redshift}, with $\alpha=-0.97\pm0.1$. This result is
in agreement to with what has been found in previous kinematical studies
(\citealt{vanderwel2008}; \citealt{newman2010}). The scatter between
different studies is considerable, with the work by
\citet{vandokkum2009a} having the largest size difference while that by
\citet{onodera2012} having the smallest. Our sample falls in between
these two extremes, i.e.,we find smaller sizes as compared to
\citet{onodera2012}, but larger effective radii than
\citet{vandokkum2009a}.

Instead of comparing galaxies sizes at fixed dynamical mass, we will
now take into account that galaxies do grow in mass
(e.g.,\citealt{patel2012}). In \citet{vandokkum2010} they find that,
for a sample selected at a constant number density, the stellar mass
evolves as
\begin{equation} log M_{n}/M_{\odot} = 11.45-0.15z
\label{eq:mass_evolution}.
\end{equation}
The number density on which this result is based, $n = 2 \times
10^{-4}$Mpc$^{-3}$, corresponds to an average mass of $\log
M_*/M_{\odot} \sim 11.2$ at $z\sim 2$, similar to our sample. Assuming
that the mass evolves as $\Delta \log M/M_{\odot} \sim 0.15z$, we will
compare effective radii for galaxies at different redshifts. For
example, a galaxy with $\log M_{dyn} /M_{\odot}= 11$ at $z\sim 2$ will
be compared with a $z\sim 0$ galaxy with $\log M_{dyn}/M_{\odot} =
11.3 $.

However, the evolution in stellar mass is determined for a complete
sample of both star forming and quiescent galaxies, while in this
paper we only look at quiescent galaxies. Therefore, we assess whether
the evolution in size at constant cumulative number density is
different for the quiescent population as compared to the full
population. This was already done for galaxies in CANDELS by
\citet{patel2012} at $n_{cum} = 1.4\times10^{-4} \rm{Mpc}^{-3}$, which
corresponds to a median mass of $\log M_*/ M_{\odot} \sim 10.9$ at
$z\sim1.8$. The sample studied here, however, has a median mass of
$\log M_*/M_{\odot} \sim 11.2$ at $z\sim1.8$, which corresponds to
$n_{cum} = 2.5\times10^{-5} \rm{Mpc}^{-3}$. Thus, we repeat the
analysis by \citet{patel2012}, but now using our CANDELS sample
(Section \ref{sec:bias}) at this lower constant cumulative number
density. Our results are similar to \citet{patel2012}, i.e.,the
quiescent population has smaller effective radii as compared to full
population, but the difference in size between the quiescent and
star-forming population is slightly smaller at $z<1.8$ as compared to
\citet{patel2012}. The difference is due to the fact that at our lower
constant cumulative number density, we find a higher fraction of
quiescent galaxies. Therefore, the effective radii of the full
population will be closer to the effective radii of the quiescent
population, as compared to \citet{patel2012}. To correct for this
difference in size, we increase the effective radii of all the
quiescent galaxies in our combined sample by the size difference of
the quiescent population as compared to the full. We correct each
galaxy individually by finding the correction factor at this
particular redshift. Below $z<1.8$, the median correction factor is
$\sim 1.05$, while at $z\sim2.1$ the correction factor is $\sim 1.6$.

Figure \ref{fig:redshift_size}c shows the evolution in size at fixed
number density as a function of redshift. Not surprisingly, the
evolution in effective radii is more extreme, as we are now comparing
$z\sim 2$ galaxies to more massive, and therefore bigger galaxies at
$z\sim0$. Using Equation \ref{eq:delta_re_redshift}, we find that
$\alpha = -1.16\pm0.1$ provides the best fit. In conclusion, assuming
that galaxies evolve in both mass and size, we find that the effective
radii have to grow by a factor $\sim 4$ from $z\sim 2$ to the present
day.

\begin{figure*} \epsscale{1.15}
\plotone{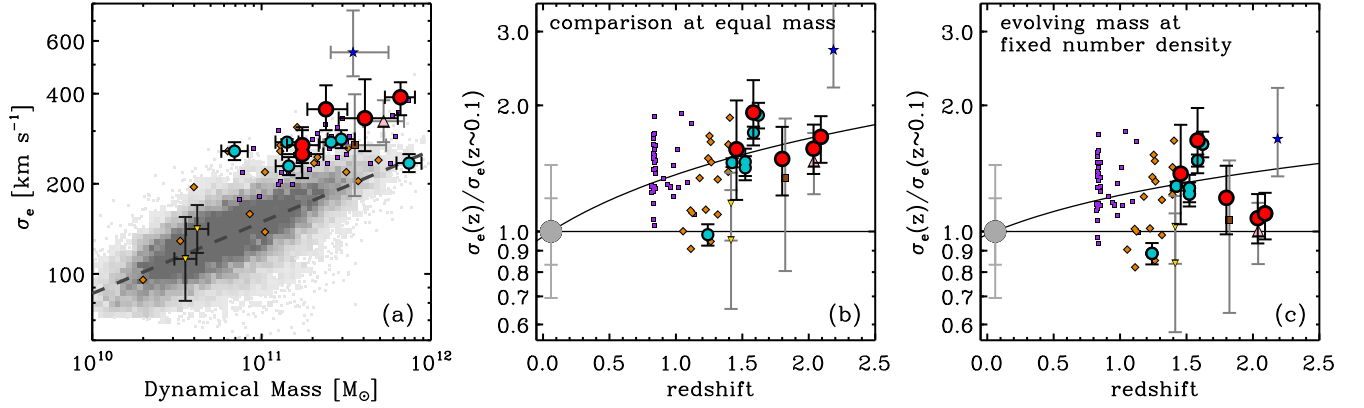}
\caption{Redshift evolution in stellar velocity dispersion within one
effective radius ($\sigma_e$). (a) $\sigma_e$ vs.
dynamical mass. At fixed dynamical mass, we find that our galaxies
have higher velocity dispersion as compared to low-redshift
galaxies. (b) Evolution of velocity dispersions at
fixed dynamical mass vs. redshift. We find the $\sigma_e \propto
(1+z)^{0.49\pm0.08}$, i.e at fixed dynamical mass the velocity
dispersion decreases by a factor of $\sim1.7$ from $z\sim2$ to the
present day. (c) Evolution of the velocity
dispersion for an evolving mass function at constant number
density. Opposite to the evolution in size, we now find a milder
evolution of $\sigma_e \propto (1+z)^{0.31\pm0.08}$.}
\label{fig:redshift_sigma}
\end{figure*}


\subsection{Evolution in Velocity Dispersion}
\label{sec:sigma_evolution}

In Figure \ref{fig:redshift_sigma}a we compare the stellar velocity
dispersion within one $r_e$ vs. the dynamical mass for both low-
and high-redshift samples. The dashed line is the parametrization of
the $\sigma_e - M_{dyn}$ relation for low$-z$ galaxies using the
following equation:

\begin{equation} \sigma_e=\sigma_c \left( \frac{M_{\rm
dyn}}{10^{11}M_{\odot}}\right)^b.
 \label{eq:mdsig}
\end{equation} We find that $\sigma_c=148.9~$km s$^{-1}$ and $b=0.24$.
Our high-redshift sample is clearly offset from low-redshift galaxies
in the SDSS, i.e.,at fixed mass they have higher velocity
dispersions. Comparison of the velocity dispersion at fixed dynamical
mass, as seen in \ref{fig:redshift_sigma}b, shows a clear evolution in
$\sigma_e$, such that velocity dispersion decreases over time. From
this figure the increase in accuracy for the velocity dispersion
measurements of this study, as compared to other studies at similar
redshift, is also clearly noticeable. Again we use the following
simple relation to quantify the amount of evolution:
\begin{equation} \sigma_e \propto (1+z)^{\alpha},
\label{eq:delta_sigma_redshift}
\end{equation}
and find that $\alpha = 0.49\pm0.08$. From $z\sim2$ to $z\sim0$ the
stellar velocity dispersions decrease by a factor $\sim 1.7$. Again,
we note that we apply a correction to the velocity dispersions in our
sample, in order to correct for the bias towards more compact galaxies
(section \ref{sec:bias}).

If we now compare low- and high-redshift galaxies using an evolving
mass function as described above, we find that the velocity dispersion
decreases less with cosmic time than when compared at fixed dynamical
mass.  Here, we also take into account that quiescent galaxies at
constant cumulative number density are smaller as compared to the full
sample, and therefore they have higher velocity dispersions. Similar
to as described above, we therefore corrected the velocity dispersions of
all galaxies in our combined sample. At $z<1.8$, velocity dispersions
on average decrease by a factor of $\sim \sqrt{1.05}$, while at
$z\sim2.1$ they decrease by a factor of $\sim \sqrt{1.6}$. If we use
equation \ref{eq:delta_sigma_redshift}, we find that
$\alpha=0.31\pm0.08$. In other words, the velocity dispersion within
one $r_e$ decreases by a factor $\sim 1.4$ from $z\sim2$ to
present-day.


\begin{figure*}
\epsscale{1.15}
\plotone{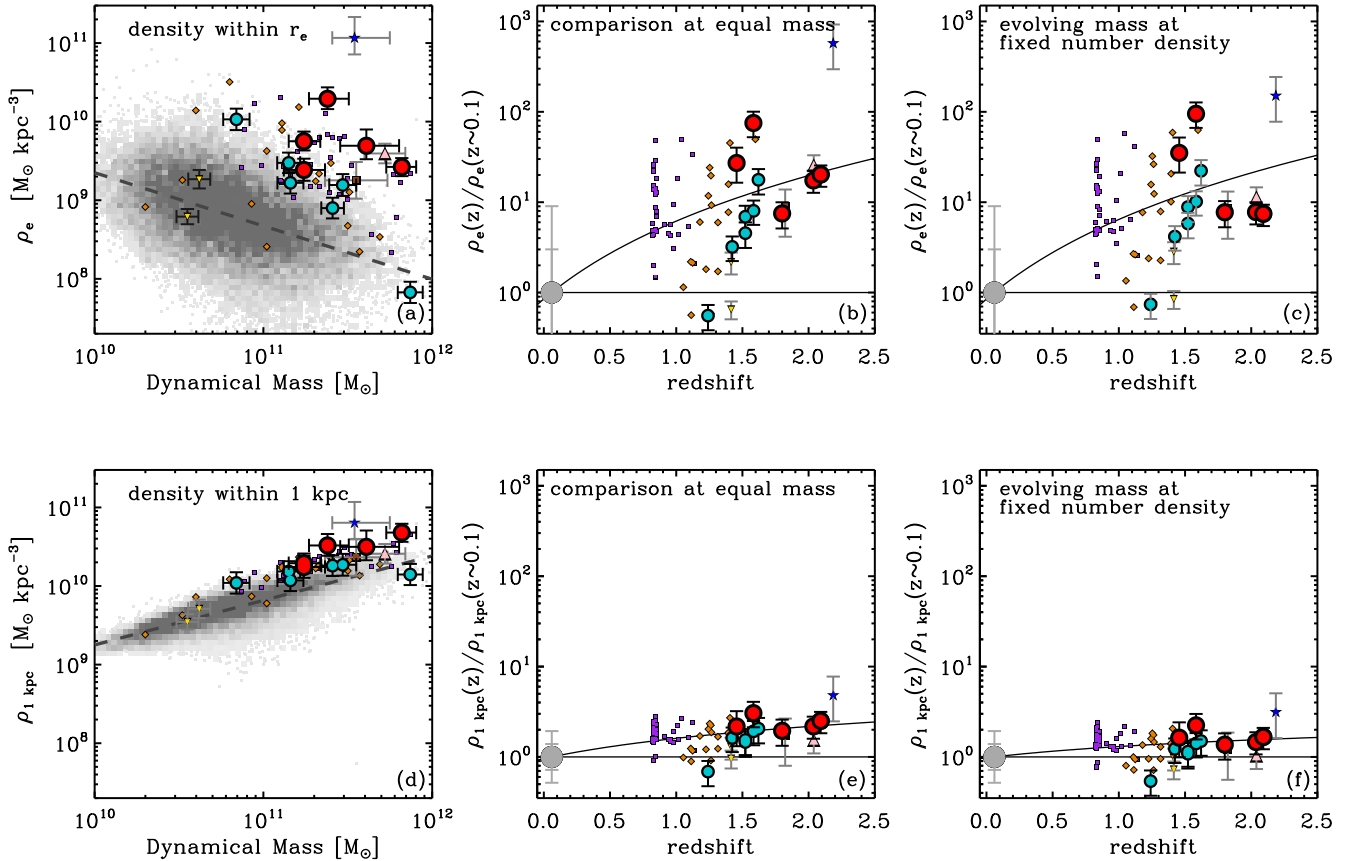}
\caption{Redshift evolution of the central and effective densities, as
calculated according to Equations \ref{eq:rho1} and \ref{eq:rhoe}. Top
row shows the results for the effective mass density, while the bottom
row shows the mass density within 1 kpc. \textit{Left panels:} the
density within $r_e$ vs. dynamical mass (a). We find
that low-redshift galaxies show a large scatter at fixed dynamical
mass. Still, our high-redshift galaxies have higher densities within
$r_e$ at fixed mass. In contrast to $\rho_e$, the density within 1 kpc
vs. dynamical mass (d) shows very little scatter, and
we find only a small difference between low- and high-redshift
galaxies. \textit{Middle panels:} evolution in the density at fixed
dynamical mass vs. redshift. We find a strong evolution for the
effective density (b) with $\rho_e \propto
(1+z)^{2.86\pm0.15}$, or a decrease by a factor of $\sim2$ from
high-redshift to $z\sim0$. For the central density, however, we only
find that $\rho_{1 kpc} \propto (1+z)^{0.74\pm0.15}$, or a decrease of
$\sim 2.3$. \textit{Right panels:} evolution of the density with an
evolving mass at constant number density. Similar to the
effective radii, we find a stronger evolution for the effective
density, i.e.,$\rho_{e} \propto (1+z)^{2.93\pm0.15}$, while the central
density requires very little to no evolution ($\rho_{1 kpc} \propto
(1+z)^{0.42\pm0.15}$).}
\label{fig:redshift_rho}
\end{figure*}

\subsection{Evolution in Mass Density}
\label{sec:density_evolution}
Next, we will focus on the central and effective mass densities using a
similar approach as described in \citet{saracco2012}. In short, using
the intrinsic S\'ersic profile we can calculate the fraction of the
luminosity that is within 1 kpc as compared to the total
luminosity. For a S\'ersic profile this ratio is given by
\citep{ciotti1991}:
\begin{equation} \frac{L_{1kpc}}{L_{tot}}
=\frac{\gamma(2n,x)}{\Gamma(2n)}.
\label{eq:lum_ratio}
\end{equation}
Here, $\Gamma (2n)$ is the complete gamma function, $\gamma (2n,x)$
the incomplete gamma function, $x=b_n (r_{1kpc} / r_{e})^{1/n}$, with
$b_n = 1.9992n-0.3271$. Using this ratio we can now calculate the
dynamical mass within one 1 kpc and within $r_e$ from the total mass:
\begin{equation} {M_{1kpc}} =\frac{L_{1kpc}}{L_{tot}} M_{dyn}.
\label{eq:mass1}
\end{equation}
Here we assume that the dynamical mass profile follows the light
profile, and furthermore that the mass-to-light ratio of the galaxy is
radially constant. The detection of small color gradients in our
galaxies indicates, however, that this is not the case, but the effect
on the derived densities is small (\citealt{saracco2012}; see also
\citealt{szomoru2012}). Finally, the densities are calculated as
follows:
\begin{equation} {\rho_{1kpc}} = \frac{M_{1kpc} }{4/3 \pi r_{1kpc}^3},
\label{eq:rho1}
\end{equation}
and
\begin{equation} {\rho_{e}} = \frac{0.5 M_{dyn}}{4/3 \pi r_e^3 }.
\label{eq:rhoe}
\end{equation}
As for $r_e$ and $\sigma_e$, we now compare the density as a function
of dynamical mass (see Figure \ref{fig:redshift_rho}). The top row
shows the results for the mass density within one effective radius,
while the bottom row compares the central density within 1 kpc. The
first thing to notice is the large scatter for low-redshift galaxies
when looking at $\rho_e$ vs. $M_{dyn}$, while $\rho_{1kpc}$ vs.
$M_{dyn}$ shows a tight relation. The density-mass relation can be
parametrized by:
\begin{equation} \rho=\rho_c \left( \frac{M_{\rm
dyn}}{10^{11}M_{\odot}}\right)^b.
 \label{eq:mrho}
\end{equation}
For the density within $r_e$ we find $\rho_{c,e} = 4.7\times10^8
$M${_\odot} $kpc$^{-3}$ and $b_e=-0.68\pm 0.15$, and for the central
density within 1 kpc $\rho_{c,1kpc} = 6.6\times10^9 $M${_\odot}
$kpc$^{-3}$ and $b_{1kpc}=0.56$.

When we compare the galaxies in our high-redshift sample to galaxies in
the SDSS, we find that they have higher densities within
$r_e$. Comparison at equal dynamical mass shows that the effective
densities are higher by a factor of $\sim50$ (Figure
\ref{fig:redshift_rho}b) for our sample. The same comparison, but now
for the central density within 1 kpc, reveals only mild evolution,
approximately a factor of $\sim3$ from $z\sim2$ to the present. When fitting
\begin{equation} \rho \propto (1+z)^{\alpha},
\label{eq:delta_rho_redshift}
\end{equation} we find that $\alpha_e = 2.86\pm0.15$, while
$\alpha_{1kpc} = 0.74\pm0.15$.

Instead of comparing galaxies at fixed mass, we again take into
consideration that galaxies evolve in mass when comparing low- and
high-redshift galaxies. Again, we correct for the fact that quiescent
galaxies at constant cumulative number density are smaller as compared
to the full sample. This time, we find that $\rho_e$ evolves even
faster as compared to the equal mass comparison $\alpha_e =
2.93\pm0.15$. The density within 1 kpc, however, requires a decrease
less than a factor of $\sim2$, with $\alpha_{1kpc} = 0.42\pm0.15$,
from $z\sim2$ to the present

\section{Discussion}
\label{sec:inside_out}
In the previous section we have found that in order for the
high-redshift galaxies in our sample to evolve into typical
present-day early-type galaxies, strong structural evolution is
required. Effective radii need to increase, and the velocity
dispersion within $r_e$ has to decrease. The density within the
effective radius has to decrease by more than an order of
magnitude. However, the central density can remain almost the same,
consistent with inside-out growth.

The dominant physical mechanism for this structural evolution is still
a subject of ongoing debate. Size growth dominated by major mergers
seems to be unlikely as it would increase the masses too much, which
would make extremely massive galaxies too common in the local
universe. As the mass and size increase at approximately at the same
rate in major mergers, the galaxies would also remain too compact for
their mass. Minor merging could offer a solution to the problem, as it
can grow a galaxy in effective radius ($r_e$) steeper than $r_e
\propto M_*$ (\citealt{villumsen1983}; \citealt{naab2009};
\citealt{bezanson2009}; \citealt{hopkins2009b} ). In this scenario,
the observed compact high-redshift galaxies may simply be the cores of
local massive early-type galaxies, which grow inside-out by accreting
(smaller) galaxies, and thus assemble a significant part of their mass
at later times (\citealt{vanderwel2009}; \citealt{oser2010}). In this
section we will examine whether dry minor merging agrees with our
findings.

From a simple estimate, based on the virial theorem,
\citet{bezanson2009} predict how the effective radii changes if a
massive galaxy undergoes a series of minor mergers. With only eight
1:10 mergers, the effective radii can grow by a factor of $\sim5$
while only having the mass increase by a factor of $\sim~2$. This is
also described by \citet{naab2009}, who state that if an initial
system undergoes a mass increase by a factor of 2 due to accretion of
very small systems, then the final radius of the system is four times
larger, the velocity dispersion is reduced by a factor of 2, and the
density is reduced by a factor of 32. This analytic prediction is
confirmed by their hydrodynamical cosmological simulation and
consistent with the observational size evolution as presented here.

Using hydrodynamic simulations of galaxy mergers,
\citet{hopkins2009a}, also find evidence for size evolution. When they
compare the effective radii of quiescent galaxies at fixed mass, they
find an evolution in size of $r_e \propto (1+z)^{-0.48}$ for galaxies
with $\log M_*/M_{\odot} = 11$, which is weaker than found by this
study and many others. \citet{oser2012} find a size evolution in their
hydro simulation, which is much stronger: $r_e \propto (1+z)^{-1.44}$,
on the high side of current observational results.

\citet{oser2012} find a similar evolution in velocity dispersion of
$\sigma \propto(1+z)^{0.44}$, to that found in this work. In contrast,
\citet{hopkins2009a}, predict that high-redshift quiescent galaxies
have roughly the same or at most a factor $\sim 1.25$ larger velocity
dispersions.

Evolution of the density is also discussed in both
\citet{bezanson2009} and \citet{hopkins2009b}. Based on photometric
data, both studies find that while the density within one effective
radius is higher at high-redshift, the central density of
high-redshift galaxies is very similar to local massive
ellipticals. From hydro simulations, \citet{naab2009} show that the
central density within 1kpc decreases by a factor of 1.5 from $z = 2$
to $z = 0$, caused by dynamical friction from the surviving cores of
the infalling systems. Similarly, \citet{oser2012} show that the central
density evolves only weakly, while the density within $r_e$ decreases
rapidly by more than an order of magnitude, in good agreement with
what we find here.  From a study of central galaxies in three
$\sim10^{13}\,$\msun galaxy groups, simulated at high resolution in
cosmological hydrodynamical simulations, \citet{feldmann2010} come to
the same conclusion. They find that the effective density of these
galaxies decreases by 1-2 orders of magnitude between $z=1.5$ and
$z=0$, while the density within 2 kpc stays roughly constant.

This is in contrast with the findings of \citet{saracco2012}, who find
no evidence for higher mass densities within one effective radius when
comparing their $z\sim1.5$ galaxies to low-redshift cluster
galaxies. Furthermore, the large scatter that they observed in the
effective density and the apparent evolution, is simply due a peculiar
analytic feature in the S\'ersic profile.

In Figure \ref{fig:redshift_delta_mass}a we show that the ratio of
$M_{*}/M_{dyn}$ appears to have evolved from $z\sim2$ to $z\sim0$.  As
compared to SDSS galaxies with $\log M_*/M_{\odot} >11$, we find that
the median $M_{*}/M_{dyn}$ is higher by $50\%$ at $z>1.5$, and that
$M_{*}/M_{dyn} \propto (1+z)^{0.17\pm 0.011}$.  However, this result
is uncertain due to the selection effects inherent in this sample and
large measurement errors in both masses. We note that this effect is
predicted by simulations; as the effective radius of a galaxy grows,
the dark matter fraction within $r_e$ will also
increase. \citet{hopkins2009a} predicts evolution by a factor of
$\sim1.25$ for galaxies with $\log M_*/M_{\odot} = 11$, with the
effect increasing with stellar mass. \citet{Hilz2012} also find a
strong evolution in the dark matter fraction in their
hydro-simulation, and predict that quiescent galaxies at $z\sim2$ have
lower dark matter fractions ($\gtrsim 80\%$). They mention that it is
mainly driven by the strong size increase, which therefore probes a
larger region that is dominated by dark matter.

\section{summary and conclusion}
\label{sec:conclusion}

In this paper, we present deep UV-NIR spectroscopy of five massive
($>10^{11}$ \msun) galaxies at $z\sim2$, using X-Shooter on the
VLT. These spectra enable us to measure stellar velocity dispersions
with higher accuracy than done before at this redshift: we triple the
sample of $z>1.5$ galaxies with well constrained ($\delta\sigma<100$
km s$^{-1}$) velocity dispersion measurements. We find that the
stellar velocity dispersions are high (290-450 km s$^{-1})$ compared
to equal-mass galaxies in the SDSS.

We combine these kinematic results with size measurements using GALFIT
on HST-WFC3 $H_{160}$ and UDS K-band imaging, and use these
measurements to derive dynamical masses.  Stellar masses are obtained
from SPS modeling on the VIS-NIR spectra in
combination with the available broadband and medium-band data. The
SPS-modeling shows that our galaxies have ages ranging from 0.5 to 2 Gyr,
and show no signs of on-going star formation.

We find good correspondence between the dynamical and stellar masses,
with the dynamical mass being higher by \mbox{$\sim15\%$}. Our results
suggest that stellar mass measurements for quiescent galaxies at
high-redshift are robust.

We complement our results with stellar kinematic results from other
studies at low and high redshift to study the structural evolution of
massive quiescent galaxies. At fixed dynamical mass, we find that the
effective radius increases by a factor of $\sim2.8$, while the
velocity dispersion decreases by a factor of $\sim1.7$ from $z\sim2$
to the present day.  Furthermore, we study how the mass density within
$r_e$ and 1 kpc evolves with time. We find a strong decrease of the
mass density within one effective radius (factor of $\sim21$), while
it only decreases mildly within 1 kpc (factor of $\sim2.3$). Instead
of comparing galaxies at fixed dynamical mass, we also use an evolving
mass limit as defined by fixed number density. By accounting for
concurrent mass growth in our comparison of high- and low-redshift
galaxy populations, we find an even stronger evolution in galaxy sizes
(factor of $\sim4$).  We find that velocity dispersion decreases less
dramatically with time, differing by only a factor of $\sim1.4$
between $z\sim2$ and $z\sim0$.  Finally, for the mass density within
$r_e$, we find a stronger evolution, but interestingly, the mass
density within 1 kpc is consistent with no evolution. This finding
implies that massive quiescent galaxies grow inside out.

We examine if our results are compatible with the current idea of inside-out growth
through dry minor mergers. Our findings are qualitatively consistent with predictions
from hydrodynamical simulations which show similar evolution in size, velocity dispersion, and
mass density within one effective radius.

Finally, we find that even though the stellar masses are consistent
with the dynamical masses, the ratio of $M_{*}/M_{dyn}$ may slightly
decrease with time. This, too, is predicted by minor merging
simulations, which show that the size growth due to minor merging will
also change the fraction of dark matter as compared to the stellar
mass within an effective radius.  This is due to the fact that the
dark matter profile is less steep than the stellar mass profile, and
thus the dark matter to stellar mass fraction increases with radius.

Despite the vastly improved accuracy of our derived dynamical masses
and stellar population parameters, the broader inferences of our study
are still limited by the small number of high-redshift galaxies with
such information. We have shown that our sample is biased towards
younger galaxies, compared to a stellar mass limited sample at
$z\sim2$, with smaller effective radii as compared to the full
population of quiescent galaxies at $z\sim2$. Only with a larger unbiased
sample of massive quiescent galaxies at high redshift can we start to
comprehend the final phase that massive galaxies go through in
becoming today's ellipticals.\\

\acknowledgments{
We thank the anonymous referee for the constructive comments which
improved the quality and readability of the paper.  We also thank
Daniel Szomoru for providing his residual-correct code and the galaxy
sizes for GOODS-S CANDELS; Andrew Newman for providing the corrected
stellar masses; Shannon Patel for the structural parameters of
galaxies in UDS CANDELS; and Ivo Labb\'e for helpful comments
regarding the SFRs from 24 $\micron$flux. It is a pleasure to
acknowledge the contribution to this work by the NMBS
collaboration. We also thank Adam Muzzin for useful
discussions.\\ This research was supported by grants from the
Netherlands Foundation for Research (NWO), the Leids Kerkhoven-Bosscha
Fonds. Support for the program HST-GO-12167.1 was provided by NASA through
a grant from the Space Telescope Science Institute.\\ This work is
based on observations taken by the CANDELS Multi-Cycle Treasury
Program with the NASA/ESA HST, which is operated by the Association of
Universities for Research in Astronomy, Inc., under NASA contract
NAS5-26555.\\ This publication also makes use of the Sloan Digital Sky
Survey (SDSS). Funding for the creation and distribution of the SDSS
Archive has been provided by the Alfred P. Sloan Foundation, the
Participating Institutions, the National Aeronautics and Space
Administration, the National Science Foundation, the U.S. Department
of Energy, the Japanese Monbukagakusho, and the Max Planck
Society. The SDSS Web site is http://www.sdss.org/. The SDSS
Participating Institutions are the University of Chicago, Fermilab,
the Institute for Advanced Study, the Japan Participation Group, Johns
Hopkins University, the Max Planck Institut fur Astronomie, the Max
Planck Institut fur Astrophysik, New Mexico State University,
Princeton University, the United States Naval Observatory, and the
University of Washington.}



\appendix

\section{Robustness of the Velocity Dispersion Measurements}
\label{sec:app_disp_test}

As it has only recently become possible to measure velocity dispersions at high redshift, the stability of these measurements has barely been tested. In this appendix, we will study the effect of fitted wavelength range, template choice, degree of the additive polynomial, S/N of the spectra, and the choice of stellar populations models.

\subsection{Dependence of the Velocity Dispersion on the Wavelength Range}
\label{subsec:app_wl}

\begin{figure*}
\epsscale{1.0}
\plotone{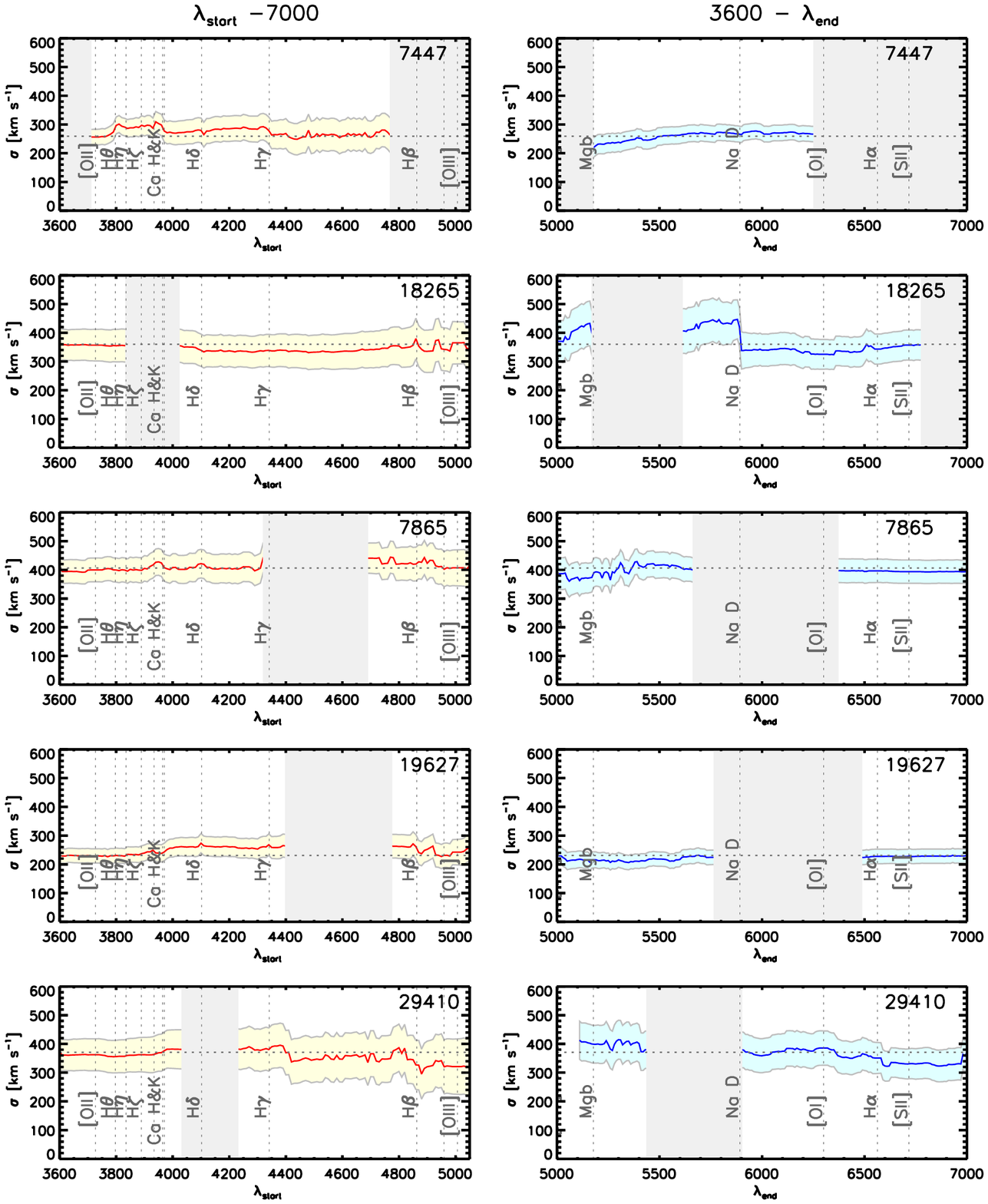}
\caption{Velocity dispersion vs. wavelength range. Left panel shows
the results for $\lambda_{start} < \lambda < 7000\AA$, while the right
panel is for $3600\AA < \lambda < \lambda_{end}$. The horizontal
shaded region indicates the 1-$\sigma$ error from pPXF, and the dashed
horizontal line indicates the velocity dispersion when fitting the
full range. Most prominent absorption lines are indicated, and the
regions affected by strong sky-lines and atmospheric absorption are
shown in gray. Overall, we find a stable solution for the velocity
dispersion while changing the wavelength range. For NMBS-C7447 and
UDS-19627, we do find a small increase in the region around the Balmer
break. For NMBS-C18265, we find that excluding Na D has a great impact
on the velocity dispersion.  }
\label{fig:wlrange}
\end{figure*}

Our sample spans a redshift range of $1.4 < z < 2.1$, which means that
different parts of the rest-frame spectra will be affected by
sky-lines and atmospheric absorption for each galaxy. This can be seen
from Figure \ref{fig:spectra_page_2}, we often lose strong absorption
features in our spectrum, which affects the region of the rest-frame
spectrum that we can fit. Here we investigate how stable the measured
velocity dispersion is as a function of the wavelength range.

For the velocity dispersion fitting in this paper, the lower
wavelength limit is set by stellar libraries and models, as no
systematic high-resolution observations exist below $3550 \AA$. The
higher wavelength limit is set by lower S/N in our spectra in the
observed K-Band. Our approach for testing the wavelength dependence of
the fit is as follows. First, we use the full-range spectrum to
determine a best-fit polynomial (1 order per 10\,000 km\,s$^{-1}$),
which is used to correct for the difference between the observed and
the template continuum. Next, we repeat the measurement with a
zeroth-order polynomial while changing the start or end
wavelength. The polynomial is not a free parameter in this fit, as
this would make it impossible to separate between the effect of the
polynomial and the wavelength range. Note also, that we use a single
template for all fits as determined from the full Visual+NIR spectrum
together with broadband and medium-band data (see Section
\ref{subsec:sps}).

\begin{figure*}
\epsscale{1.05}
\plotone{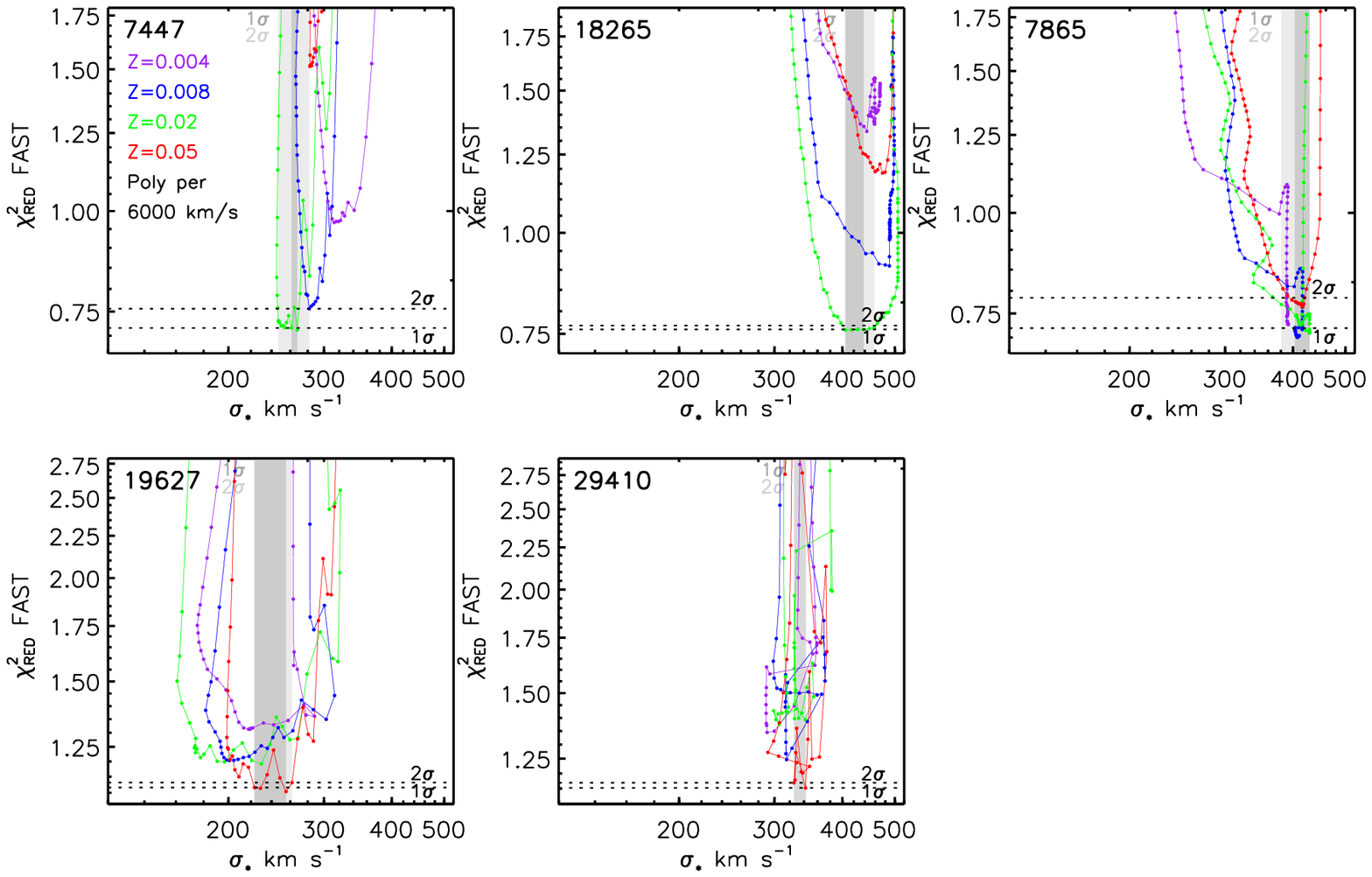}
\caption{Reduced $\chi^2$ from SPS-modeling vs. velocity dispersion
for different templates. Different points indicate different template
ages, while different colors show different metallicities. All fits
are done on the full-wavelength range and with one-order per
6000km\,s$^{-1}$ for the additive polynomial. The narrower the
horizontal distribution is, the more stable the velocity dispersion
is. For most sources, we find a stable solution for the velocity
dispersion by using different templates, except for UDS-19627, which
shows a large range in velocity dispersions. }
\label{fig:disp_chi}
\end{figure*}
\begin{figure*}
\epsscale{1.05}
\plotone{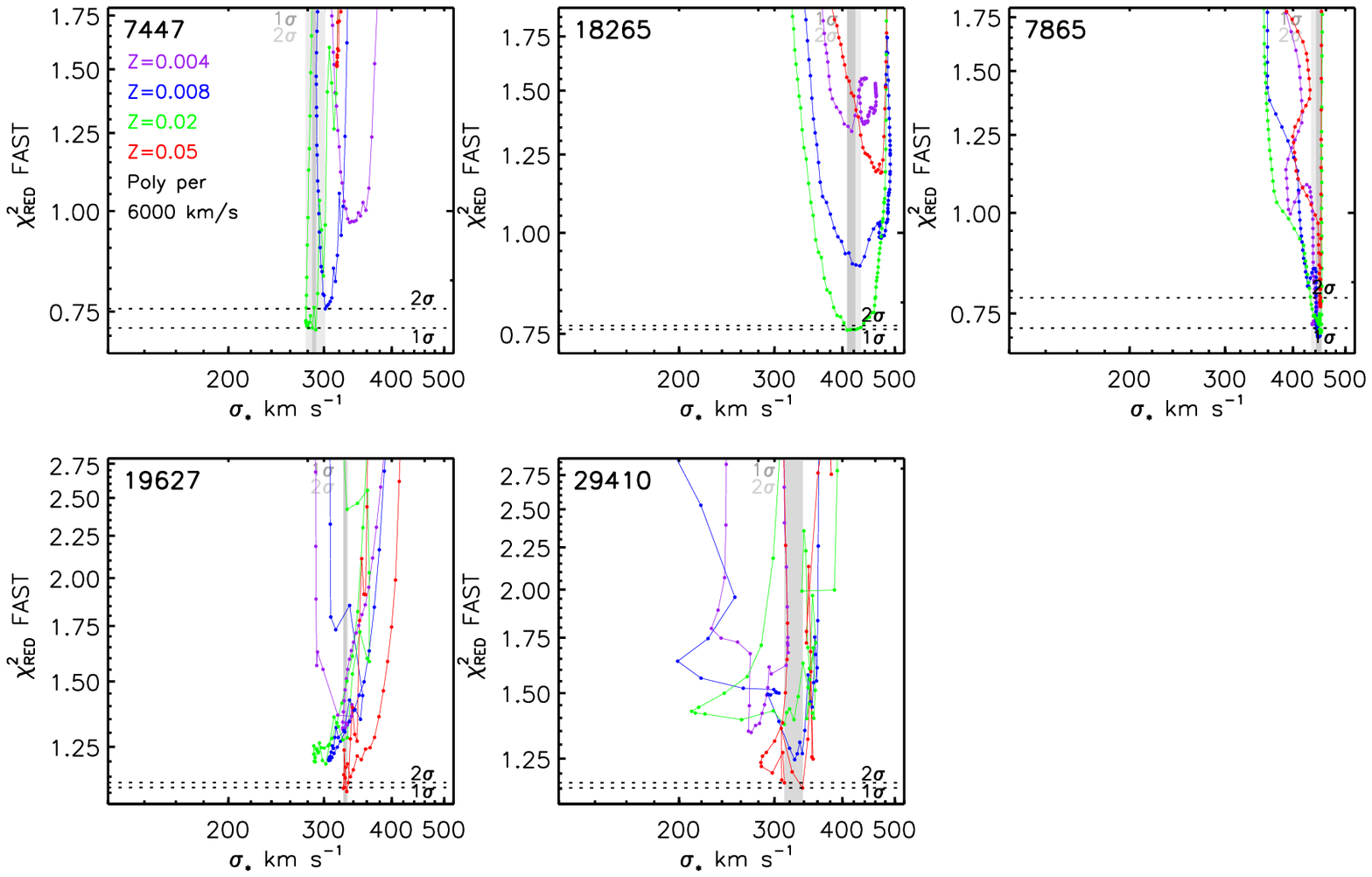}
\caption{Reduced $\chi^2$ vs. velocity dispersion for different
templates, similar to Figure \ref{fig:disp_chi} but now excluding the
Balmer break ($\lambda > 4020\AA$). Whereas with the full wavelength
range UDS 19627 showed a large variation in the velocity dispersion as
a function of template, this time we do find a stable solution, as the
distribution is narrower. The other sources also, show a decrease in
the uncertainty due to different templates, except for
UDS-29410. However, this is due to lower S/N of this spectrum and the
shorter fitted wavelength range as before.
}
\label{fig:disp_chi_gt4020}
\end{figure*}

Figure \ref{fig:wlrange} shows the results for the different
sources. The left panel shows the result where we change the starting
wavelength, i.e.,the wavelength range is $\lambda_{start} < \lambda <
7000\AA$, whereas the panels on the right show the effect of changing
the end wavelength, $3600\AA < \lambda < \lambda_{end}$. The first
thing to notice is that the measured dispersions are remarkably
stable, even when most of the absorption lines have been excluded from
the fit. The two galaxies with the youngest ages and strong Balmer
absorption lines (NMBS-C7447 and UDS-19627) show a change in the
velocity dispersion in the region of the Balmer break. NMBS-C7447
shows an increase for $\lambda_{start} > 3800\AA$, but decreases after
Ca H\&K have been removed from the fit. UDS-19627 shows an increase of
50 km\,s$^{-1}$ when the Balmer break is excluded. When we reduce the
red part of the spectrum (Figure \ref{fig:wlrange}, right panels), we
also find a stable fit, except for NMBS-C18265. After excluding Na D
from the fit, the velocity dispersion increases by a $\sim$100
km\,s$^{-1}$. We think this is because NMBS-C18265 has a more evolved
stellar populations than say for example NMBS-C7447. With the Ca H\&K
lines masked out due to atmospheric absorption, Na D is one of the
strongest lines in the spectrum, and its exclusion could explain the
sudden change in the measured dispersion.

To summarize, for most galaxies we find only a mild dependence on the
wavelength range that is used in the fit. We do find that with
decreasing wavelength range, the random error increases. Finally, even
in the absence of strong absorption features like Ca H\&K, we find
similar velocity dispersions as compared to the full range fit. Due to
template mismatch, we only fit $ \lambda>4020 \AA$ in our final
results, as is explained in the next section.

\subsection{Dependence of the Velocity Dispersion on the Template Choice}
\label{subsec:app_wl}

Next, we study how different templates may influence the measured
velocity dispersion. We use a sample of BC03 $\tau$-models, as
presented in Section \ref{subsec:sps}. In particular, we are
interested in the effect of template age and metallicity. In Figure
\ref{fig:disp_chi} we show the reduced $\chi^2$ from the SPS-modeling,
vs. the velocity dispersions measured using this template. The
different points represent different ages of the templates, with the
minimum $\chi^2$ corresponding to the best-fit age as listed in Table
\ref{tab:stellar_prop}. The different colors indicate the different
metallicities. We show the $\chi^2$ from the SPS-modeling instead of
the $\chi^2$ from the dispersion fit, as the former is derived from
the full Visual and NIR spectrum plus all the broadband and medium-band
photometric data. This large wavelength range yields better
constraints for the stellar population, thus larger relative ranges in
$\chi^2$ as compared to the $\chi^2$ from the dispersion fit. Also, as
we add high-order additive polynomials to the templates from fitting
the velocity dispersion (in this case one order per 6\,000
km\,s$^{-1}$), the effect of different template-ages is mostly washed
out, and we get a small relative $\chi^2$.

\begin{figure*}
\epsscale{1.15}
\plotone{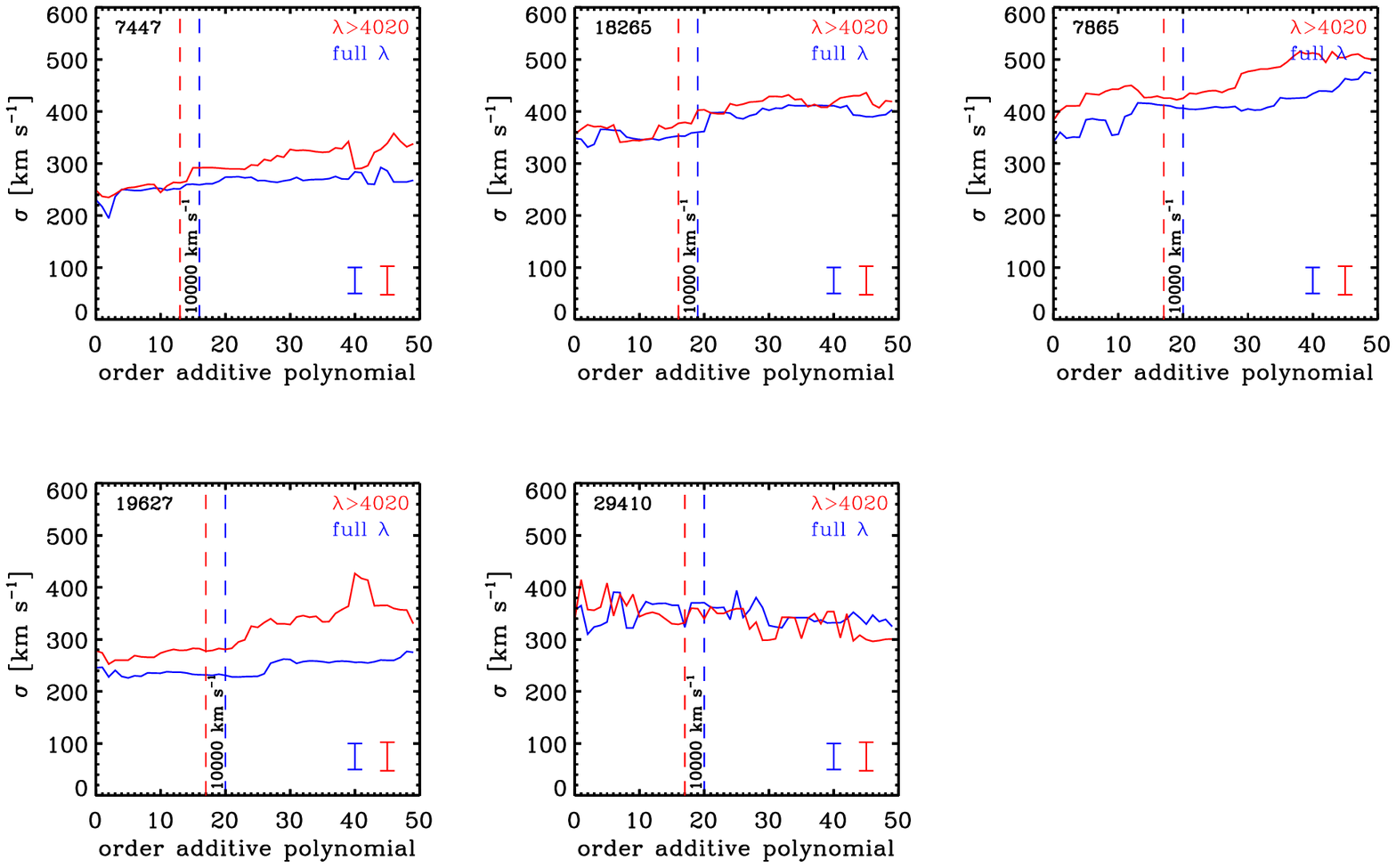}
\caption{Velocity dispersion as measured with different order additive
polynomials. The full-wavelength range fit is shown in blue, while the
fit with $\lambda > 4020\,\AA$ is shown in red. We find an increase of
the velocity dispersion with increasing order. The fit is most stable
when using a polynomial with order between 10 and 30. The vertical
dashed lines shows the order that corresponds to 10000 km\,s$^{-1}$,
the polynomial that we use for obtaining science results.}
\label{fig:polycheck}
\end{figure*}

In Figure \ref{fig:disp_chi} we see that for most galaxies the
velocity dispersion for templates allowed within $2\sigma$ gives
consistent results. Different metallicities do give different velocity
dispersions at their minimum $\chi^2$, but this is a mere reflection
of the age-metallicity degeneracy. Different metallicities have
different best-fitting ages, which in turn give different velocity
dispersion. At a $1-\sigma$ level, we only have a handful of
best-fitting templates, for which we obtain similar velocity
dispersions. UDS-19627 is the exception, which shows a large
dependency of the measured velocity dispersion as a function of
template age within the $1-\sigma$ allowed range. At the $1-\sigma$
level, we find a range of $\sim30$km\,s$^{-1}$, due to template
uncertainty, while the random error is one of the lowest, only $\sim
30$km\,s$^{-1}$. Even though templates with $\chi^2$ of 1.20 or higher
are statistically considered a bad fit, the large range in the
velocity dispersion is worrying. If we could not have constrained the
best-fitting template from the full range spectrum and broadband
data, the measured velocity dispersion would be highly uncertain.

However, if we exclude the Balmer break from the velocity dispersion
fit, the dependency on template age almost completely disappears
(Figure \ref{fig:disp_chi_gt4020}). UDS-19627 shows the most dramatic
change, where we suddenly see a tight range of best-fitting velocity
dispersions. We do note that the velocity dispersion has increased, as
was already shown in Figure \ref{fig:wlrange}. UDS-29410 appears to
have a slightly higher template uncertainty if we only fit for
$\lambda > 4020\,\AA$, but this is driven by the lower S/N of the
spectrum as we now use a shorter wavelength range.

To conclude, we find a systematic uncertainty due to templates with
different ages. This is caused by the Balmer break, present in the
relatively young galaxies in our sample. By only fitting for $\lambda
> 4020\,\AA$, the uncertainty due to template mismatch almost completely
disappears. In that case, templates with different metallicities do
give different velocity dispersions, but this is most likely caused by
the underlying age-metallicity degeneracy.

\subsection{Dependence of the Velocity Dispersion on the Order of the
Additive Polynomial} In order to correct for stellar continuum
emission differences between the observed galaxy spectrum and the
template, we use an additive Legendre polynomial. If we would not
apply such a correction, the fitting routine could try to correct for
this discrepancy by changing the velocity dispersion. Values that are
typically used in the literature vary from 5000 to 15000 km\,s$^{-1}$
per order. We do not use multiplicative Legendre polynomials, because
the S/N of the spectra are too low, and it would add another degree of
uncertainty to the fit. Here, we test the influence of the additive
polynomial to our measured velocity dispersions. Again we use the
best-fit $\tau$-model as a template while varying the additive
polynomials from 0 to 50. We fit both the full-wavelength range and
the wavelength range with $\lambda > 4020\,\AA$.

Figure \ref{fig:polycheck} shows the results, with the blue line
representing the full-wavelength fit, while the red line shows the
results for $\lambda > 4020\,\AA$. The vertical dashed line indicates
the polynomial with one order per 10000 km\,s$^{-1}$. Overall we find
that by increasing the additive polynomial, the measured velocity
dispersion increases. In general we find that between polynomials of
10 and 30, the smallest increase in the velocity dispersion occurs,
and this appears to be the most stable region. For this reason, we use
one order per 10000 km\,s$^{-1}$ for our science results.

\begin{figure*}
\epsscale{1.15}
\plotone{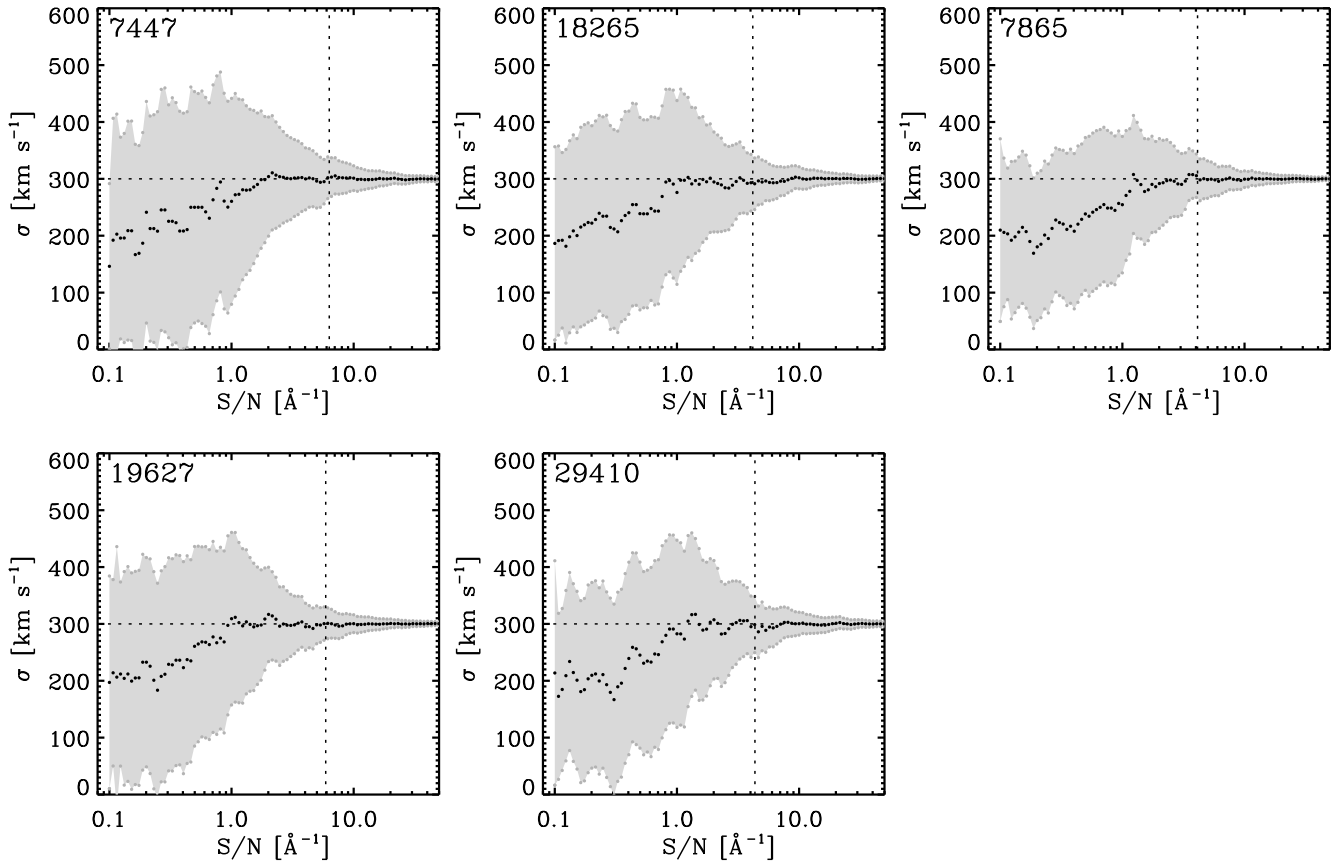}
\caption{Comparison of the recovered velocity dispersion vs. the
S/N of simulated spectra. The vertical dashed line shows the original
S/N of the spectrum, and the horizontal dashed line shows the input
velocity dispersion of 300 km s$^{-1}$. The black dots show the median
of a 100 simulations for each S/N, while the gray area shows the
median absolute deviation from these 100 simulations. We find no
systematic offset from the input velocity dispersion for S/N $>
1\AA^{-1}$, while below this S/N we find that the recovered velocity
dispersion is systematically lower.
}
\label{fig:disp_sn}
\end{figure*}

\subsection{Systematic Errors on the Velocity Dispersion at Low Signal-to-Noise}
While it is clear that a decrease in the S/N of the spectra will
increase the random errors on the velocity dispersion, it could also
cause systematic offsets (e.g.,see \citealt{franx1989}). A concern
would be that low S/N would lead to an overestimate of the velocity
dispersions.  Therefore, in this section we test if there are
systematic effects when changing the S/N for each source. The S/N we
refer to in this section is derived in the region in which we also fit
the velocity dispersion, i.e.,$4020 < \lambda_{rest-frame} [\AA] <
7000$.

For each source, we calculated the effect of S/N in a similar fashion
as we derived the uncertainty on the velocity dispersion in Section
\ref{subsec:dispersions}. First, we subtracted the best-fit model from
the spectrum. This residual now corresponds to the noise for this
particular S/N of this source. We multiply this residual by a certain
factor to obtain a new (higher or lower) S/N as compared to our
reference S/N. These residuals are then randomly rearranged in
wavelength space and added to the best-fit template that has been
convolved to a velocity dispersion of 300 km s$^{-1}$. We then
determined the velocity dispersion of simulated spectra with this S/N,
and repeat this shuffling 100 times. Lastly, we repeat this
procedure for a large range in S/N from 0.1 to 100 $\AA^{-1}$.  We
note that uncertainties due to template mismatch are not included in
this analysis.

In Figure \ref{fig:disp_sn} we show the results, with the different
values of the S/N on the horizontal axis. We see that increasing the
S/N will also decrease the random error in the velocity
dispersions. On average, we find that in order to reach a uncertainty
of less than $10\%$ on the velocity dispersion measurement, you need an
S/N of 8$\AA^{-1}$ or higher. NMBS-COS 7865 shows the lowest random
error at fixed S/N, which is older stellar population of the
galaxy. No systematic offset in the velocity dispersion is found for an
S/N of $\gtrsim 1 \AA^{-1}$. Below this value, however, we find that
the velocity dispersion is systematically underestimated as the S/N
decreases, though the offset is still smaller than the random
errors. We note that \citet{franx1989} already presented simple
arguments for predicting the systematic error. They show that the
systematic offset of the velocity dispersion can be derived from error
in the measured velocity, and goes as $\Delta \sigma \propto 0.5 dv^2
/ \sigma$.  We verified from these simulations that this is indeed the
case. From this analytic approximation, we furthermore conclude that
the systematic uncertainties of our measured velocity dispersion are
small.

\subsection{Dependence of the Velocity Dispersion on the SPS Models and Template Construction}
Here, we test how our velocity dispersions would change if we make
different choices for the SPS-model, and test the difference between a
single $\tau$-model and a template constructed from different
combinations of singe stellar population (SSP) models.

\begin{figure*}
\epsscale{1.15}
\plotone{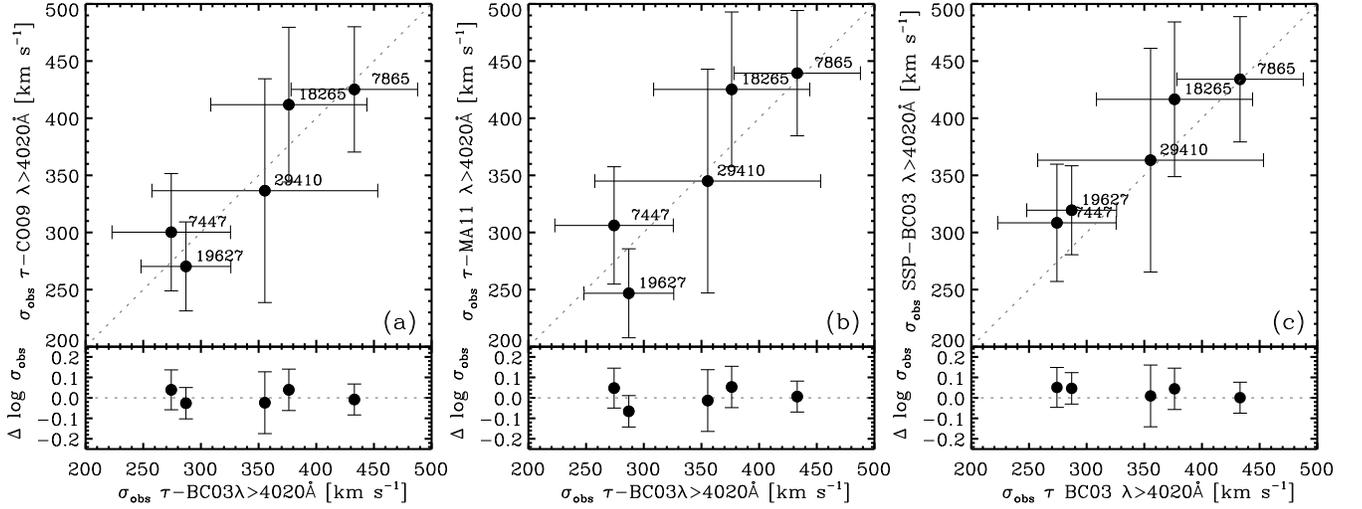}
\caption{Comparison of the velocity dispersion when using different
methods. \textit{Left panel:} velocity dispersion as derived with the
models of BC03 and \citet{conroy2009}. Even though the models are
based on different stellar libraries with different resolution, the
velocity dispersions measured with these models are
consistent. \textit{Middle panel} the same as the left panel but now
using BC03 and \citet{maraston2011} for comparison. In general, we
find consistent results within the errors, but UDS19627 has a lower
velocity dispersion when we use the MA11 models. \textit{Right panel:}
velocity dispersion as derived with optimal template construction
vs. the best-fitting $\tau$-model. The optimal template is
constructed from BC03 SSP models with a full range in age and
metallicity. Dispersions from the optimal template construction are
slightly higher, but well within the errors.  }
\label{fig:disp_comp}
\end{figure*}

The left and middle panel of Figure \ref{fig:disp_comp} show what
would happen if we would choose the Flexible-SPS models by
\citeauthor{conroy2009} (\citeyear{conroy2009}, CO09) or the models by \citeauthor{maraston2011} (\citeyear{maraston2011}, MA11). These models are based on a different stellar library with
slightly higher resolution as compared to BC03, i.e.,MILES
\citep{miles2006} vs. STELIB \citep{stelib2003}. If a systematic
uncertainty in the measured velocity dispersion is present, for
example due to the resolution or details that go into the SPS-models,
it would show up in this comparison. We determined a best-fit $\tau$
model using the CO09 and MA11 models in exactly the same way as was
done for the BC03 models (see Section \ref{subsec:sps}). When
comparing the velocity dispersions derived by using the BC03 and CO09
$\tau$-models, we do not find any significant difference (left panel
of Figure \ref{fig:disp_comp}). In the middle panel we compare MA11
and BC03. Here too, we find a good correspondence. For UDS19627 the
MA11 models give a lower velocity dispersion, though still within the
$1-\sigma$ error. These results confirm that our measurements are
stable against the choice of SPS-model and template with different
resolution.

Finally, in Figure \ref{fig:disp_comp} (right panel) we compare the
velocity dispersion from the best-fit $\tau$-model, vs. an optimal
template constructed by pPXF. This optimal template is built from a
full spectral library from BC03 SSP
models, with full range in age and metallicity. We note that the
optimal template construction only uses the wavelength regime provided
in the dispersion fit ($3600\,\AA < \lambda < 7000\,\AA$) , and does
not take into account the effects of dust. The velocity dispersions
from the optimal templates are slightly higher as compared to the
single $\tau$-model, although it is well below the random
error. Interestingly, the galaxy with the largest dust-contribution,
NMBS-C18265, also shows the largest difference between the two
different fitting techniques. In this paper, we choose to use the
best-fitting $\tau-model$ as the dispersion template, as this is the
best representation of the stellar population. As the stellar mass is
also based on this $\tau$-model, we use the same stellar population
when comparing the stellar to the dynamical mass.

\section{Aperture Corrections for Velocity Dispersion Measurements}
\label{sec:app_aper_corr}

Here, we investigate the effects of different apertures and extraction
methods on the observed velocity dispersion. The standard approach is
to use a power-law to scale the observed velocity dispersion, measured
within a certain $r_{aper}$, to the velocity dispersion as if measured
within $r_{e/8}$, using the following expression by
\citet{jorgensen1995}:

\begin{eqnarray}
\frac{\sigma_{ap}}{\sigma_{e/8}} = \left( \frac{r_{aper}}{r_{e}/8} \right) ^{-0.04}.
\label{Jorgenson}
\end{eqnarray}
\begin{figure*}[!t]
\epsscale{0.6}
\plotone{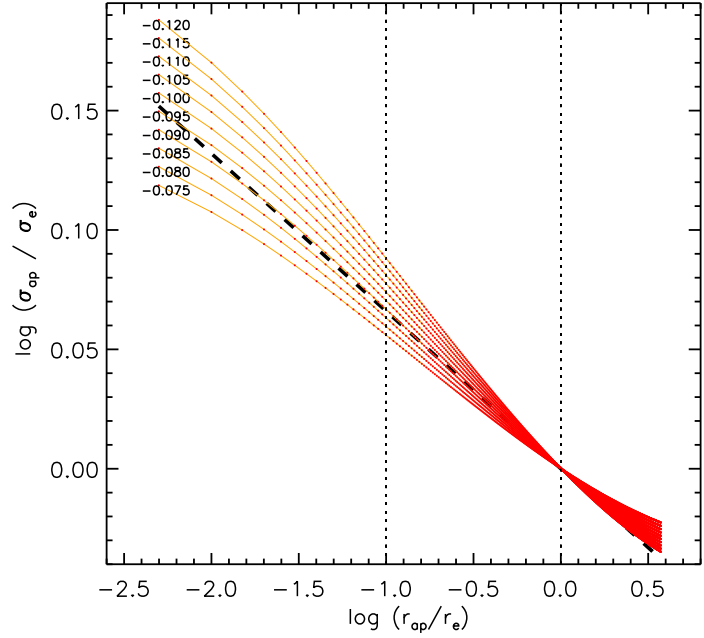}
\caption{Comparison of the observed vs. the analytic kinematic
profile. The black dashed line is the observed relation from
\citet{cappellari2006}, and is valid between  for $-1.0 <
\log (r_{\rm{aper}} / r_{e} ) < 0$ using $d=-0.089$, shown as the vertical dashed
lines. For different values of $d$ in Equation \ref{eq:sigma_profile},
we evaluated Equation \ref{eq:sigma_aper} which is shown as the
different red lines. Between $-1.0 < \log (r_{\rm{ape}r} / r_{e} ) <
0$, we find the best match between the observed and analytic kinematic
profile when using $d=-0.089$ }
\label{fig:find_power}
\end{figure*}
Or using a more recent result from the SAURON survey \citep{cappellari2006}:
\begin{eqnarray}
\frac{\sigma_{ap}}{\sigma_{e}} = \left( \frac{r_{aper}}{r_{e}} \right) ^{-0.066\pm0.035}.
\label{Cappellari et al.}
\end{eqnarray}
In the case of \citet{jorgensen1995}, the velocity dispersions have
been measured in the range of $-1.5 < \log (r_{\rm{aper}} / r_{e} ) <
-0.5 $, and for \citet{cappellari2006} for $-1.0 < \log (r_{\rm{aper}}
/ r_{e} ) < 0$. In this work however, we are outside this range with
$\log (r_{aper} / r_{e} ) > 0.5$, that is $r_e $ being much smaller
than $r_{\rm{aper}}$. Also, the galaxies in our sample have effective
radii much smaller than the FWHM of the PSF, which is why the standard
approach most likely will not be valid.

We analyze this problem in two steps. Firstly, we use an analytic
description for the kinematic profile and match this to the observed
relation as found by \citet{cappellari2006}. Secondly, using our model
we study the behavior of the observed velocity dispersion for
different apertures (both circular and non-circular) and different
FWHM for the PSF. Our reference model will be a circular aperture with
size $r_e$, without the effects of seeing.


A good description of the kinematic profile for early galaxies is given by (see e.g., \citealt{treu1999};\citealt{bertin2002}) :
\begin{eqnarray}
\sigma(r) =\left(  \frac{r}{r_{e}}\right) ^{d} \times \sigma_{cst},
\label{eq:sigma_profile}
\end{eqnarray}
with $-0.1 < d < 0$. The observed kinematic profile within a circular aperture will be a projection of $\sigma(r)^2$ and the galaxies intensity profile :
\begin{eqnarray}
\sigma^2(r_{aper}) =\frac {\int_{0}^{r_{aper}}  \sigma^2(r) I_{gal}(r) 2\pi r dr } {\int_{0}^{r_{aper}}  I_{gal}(r) 2\pi r dr } .
\label{eq:sigma_aper}
\end{eqnarray}
Here, $I_{gal}(r)$ is the S\'ersic Profile:
\begin{eqnarray}
I(r)=I_e \exp \left\lbrace -b_n \left[ \left( \frac{r}{r_e}\right)^{1/n} -1  \right] \right\rbrace 
\label{sersic_profile}
\end{eqnarray}
To avoid numerical issues in the center, we approximate Equation \ref{eq:sigma_profile} and Equation \ref{sersic_profile} by:
\begin{eqnarray}
\sigma(r) =\left(  \frac{r+r_{core}}{r_{e}}\right) ^{d} \times \sigma_{cst}, \\
I(r)=I_e \exp \left\lbrace -b_n \left[ \left( \frac{r+r_{core}}{r_e}\right)^{1/n} -1  \right] \right\rbrace 
\label{approx}
\end{eqnarray}
where $r_{core}$ is chosen to be $1/30r_{e}$. We can estimate the
power $d$ by evaluating Equation \ref{eq:sigma_aper} for different values
of $d$ and comparing the results to Equation 1 from
\citet{cappellari2006}. Figure \ref{fig:find_power} shows the
\citet{cappellari2006} relation (dashed black line) and our model with
different values for $d$, normalized to $r_e$ (red lines). We find a
best fitting value for $d=-0.089$ in the region of ($-1.0 <\ log
(r_{aper}/r_{e}) < 0$. Notice that for $\log (r_{aper}/r_{e}) > 0.0$
our model deviates from the simple power law in
Equation \ref{eq:sigma_profile}.

\begin{figure*}[!t]
\epsscale{1.1}
\plottwo{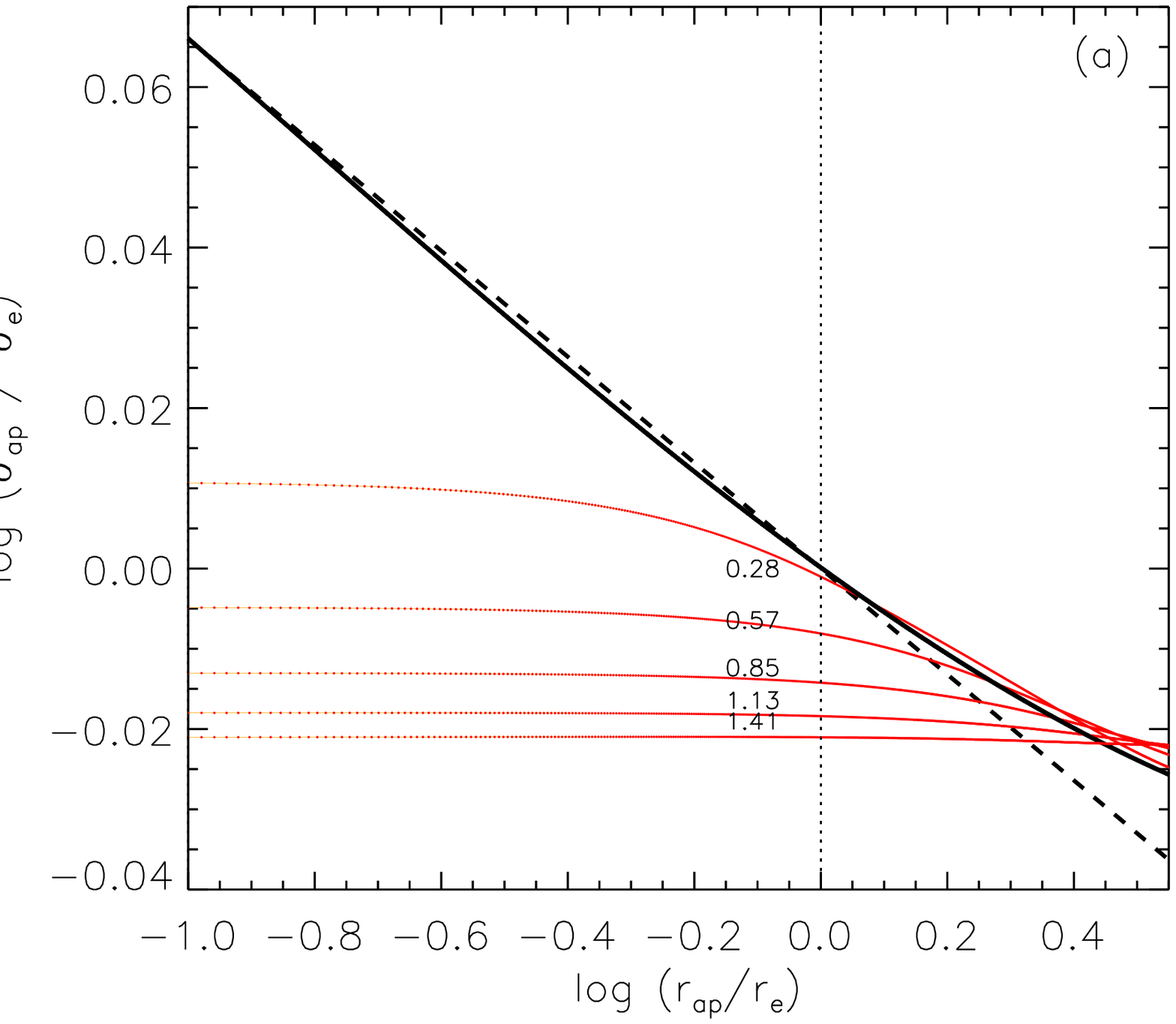}{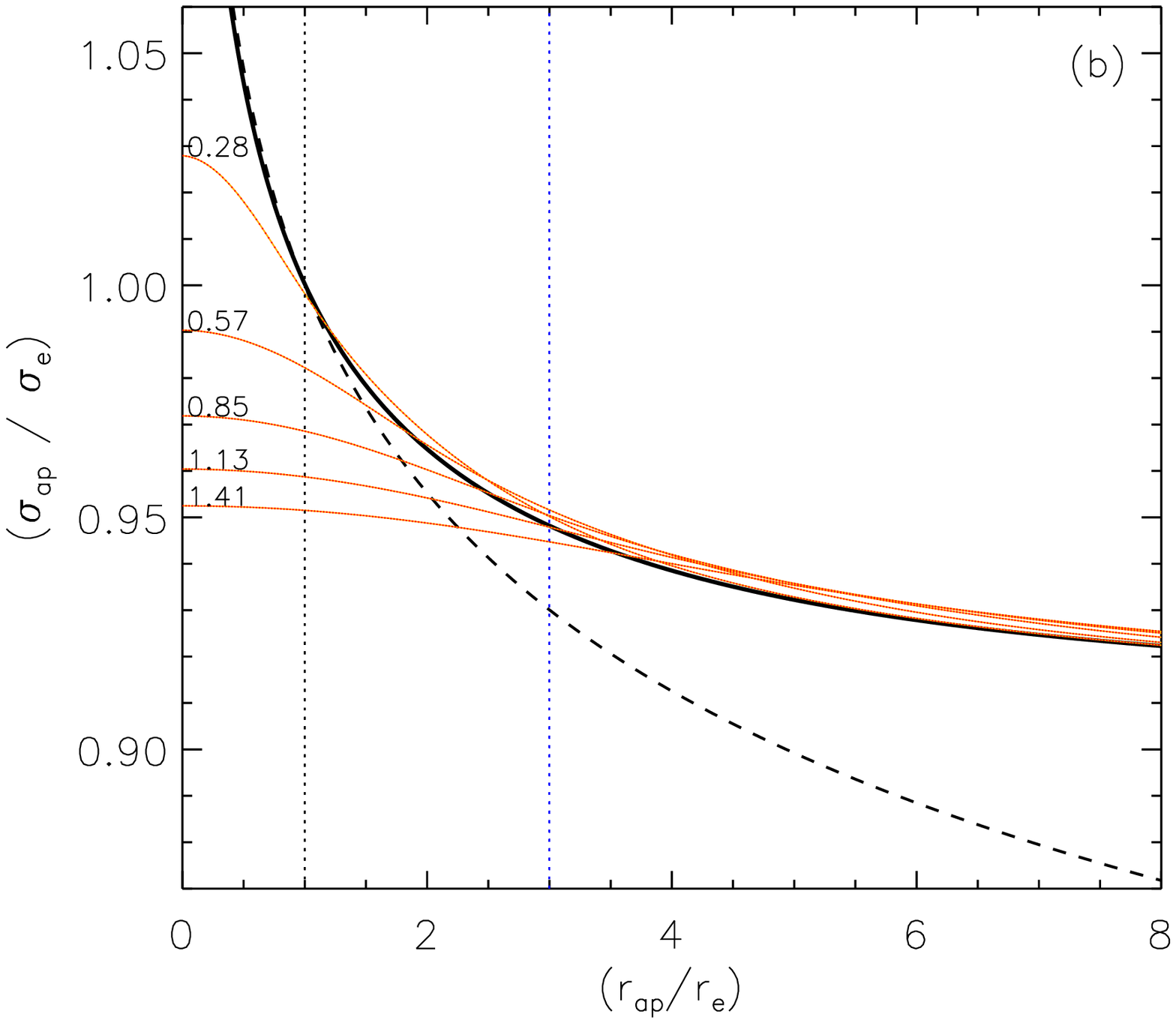}
\caption{Comparison of the observed vs. the analytic kinematic
profile, similar to Figure \ref{fig:find_power} but now including the
effect of seeing. Panel (a) shows the kinematic in log-log space,
while the panel (b) is in linear units.  The black dashed line is the
observed relation from \citet{cappellari2006}, while the solid black
line shows our reference model (no PSF). The red lines show different
assumption for the seeing, where the numbers indicate the FWHM of the
PSF in arcseconds. The vertical black dotted line shows where $r_{ap} =
r_e$, and the vertical blue dotted line shows a typical
$r_{aper}/r_{e}$ for our observations. With increasing FWHM, we find
that the kinematic profiles becomes less steep for the inner part, as
is expected from the convolution with the PSF. In the outer part of
the kinematic profile, we see very little change as compared to our
reference model.}
\label{fig:psf_effect_circ_aper}
\end{figure*}


Now that we have found the intrinsic kinematic profile, we can explore
the influence of the PSF on the observed dispersion. The PSF is
modeled using a combination of two Gaussians, where $\sigma_1
=2\sigma_2$, and both Gaussians having equal flux. This PSF is then
convolved with the kinematic and intensity profiles:

\begin{eqnarray}
\sigma^2(r_{aper}) =\frac {\int_0^{r_{aper}}\left\lbrace \left[  \sigma^2(r) I_{gal}(r) \right] \otimes PSF  \right\rbrace~2 \pi rdr} 
										{\int_0^{r_{aper}}\left\lbrace\left[  I_{gal}(r)\right] \otimes PSF  \right\rbrace~2 \pi rdr } .
\label{eq:sigma_aper_sum_psf}
\end{eqnarray}
Again we compare the observed kinematic profile from
\citet{cappellari2006} and our analytic results from Equation
\ref{eq:sigma_aper_sum_psf}, which includes the effect of the PSF, and
are shown in Figure \ref{fig:psf_effect_circ_aper}. If we compare the
black line, which is our reference model, to the red lines, which
include the effect of the PSF, we see that the inner profile ($\log
(r_{aper}/r_{e}) < 0.5$) is mostly affected and the kinematic profile
becomes less steep. This is also particularly clear from Figure
\ref{fig:psf_effect_circ_aper}b where we see that very little changes for $r_{aper}/r_{e}
> 4 $ as compared to the reference model.

\begin{figure*}[!t]
\epsscale{1.15}
\plotone{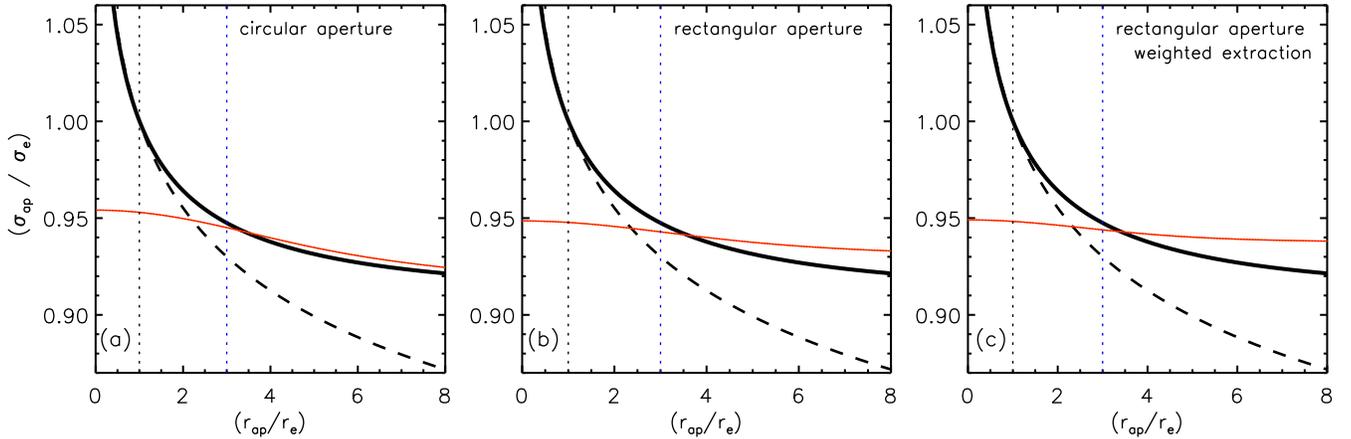}
\caption{Observed kinematic profile in different apertures: circular,
rectangular, and rectangular including a weight function according to
Equation \ref{eq:sigma_aper_sum_weight_psf}. The dashed line is the
\citet{cappellari2006} relation, the solid line is our reference model
(circular aperture, no PSF), and the red line is the modeled kinematic
including the effect of the PSF. The vertical blue dashed line shows a
typical $r_{aper}/ r_e$ for our observations.  We note that $r_{\rm
aper}$ is in the direction along the slit.}
\label{fig:psf_effect_aper}
\end{figure*}

Instead of using a circular aperture, we now consider the case of a
rectangular aperture, similar to what is used in the spectra from
X-Shooter. Equation \ref{eq:sigma_aper_sum_psf} can be modified to include
a weight function $g(y)$ which is commonly used in optimized
extraction. The integral in Equation \ref{eq:sigma_aper_sum_weight_psf} is
also replaced by a Riemann sum. In the $x$ direction (slit width) the
aperture size is always the same, i.e $0''.9$ for X-Shooter spectra,
but the $y$ direction is now parallel to $r_{aper}$:

\begin{eqnarray}
\sigma^2(x_{aper},y_{aper}) =\frac {\sum_{0}^{x_{aper}}\sum_{0}^{y_{aper}} \left\lbrace  \left[  \sigma^2(x,y) I_{gal}(x,y) \right] \otimes PSF \right\rbrace g(y)~ \Delta x \Delta y }
{\sum_{0}^{x_{aper}}\sum_{0}^{y_{aper}} \left\lbrace  \left[  I_{gal}(x,y) \right]  \otimes PSF \right\rbrace  g(y) ~ \Delta x \Delta y } .
\label{eq:sigma_aper_sum_weight_psf}
\end{eqnarray}
Figure \ref{fig:psf_effect_aper} shows the difference between using a circular aperture, a rectangular aperture without a weighting function, and a rectangular aperture including a weighting function. The slit width that was used is 0.9", with the FWHM of the PSF also being 0.9". The spectrum was extracted with $r_{aper}=0''.45$. We see that the correction is slightly higher for the rectangular aperture as compared to the circular aperture at $r_{aper} / r_e =2.25$. The behavior at $r_{aper} / r_e > 3 $ is very different for the three different cases. When using a rectangular aperture with optimized extraction, the observed profile is flatter as compared to the other models. The corrections are on average between 3$\%$ and 5$\%$.
From Figure \ref{fig:psf_effect_aper}, it is clear that by using a
simple power law for the aperture correction, we would overestimate
the corrections by a large fraction. Furthermore, we have shown that
it is vital to use a suitable aperture and include the effects of
seeing, especially when $r_{\rm aper} / r_e> 2$. The corrections
applied to our final results are derived for the rectangular aperture,
including a weighting function.

\vspace{1cm}



%

%


\clearpage

\end{document}